\documentclass[11pt,a4paper]{article}
\pdfoutput=1
\usepackage{graphicx}
\usepackage{jheppub}

\usepackage{amsmath,amssymb,mathtools}
\usepackage{nccmath}
\usepackage[compatibility=false]{caption}
\usepackage{subcaption}
\usepackage{booktabs}
\usepackage{multirow}

\def\beq{\begin{equation}}
\def\eeq{\end{equation}}
\def\bea{\begin{eqnarray}}
\def\eea{\end{eqnarray}}
\def\beu{\begin{quotation}}
\def\eeu{\end{quotation}}

\newcommand{\ud}{\,\mathrm{d}}
\def\Lp{p\cdot p}
\def\Lmax{L_{max}}

\def\Minkowski{\mathbb{M}}

\begin{document}

\title{Generalized Causal Set d'Alembertians}

\author[a,b]{Siavash Aslanbeigi,}
\author[a,b]{Mehdi Saravani}
\author[a,b,c,d]{and Rafael D. Sorkin}
\affiliation[a]{Perimeter Institute for Theoretical Physics, 31 Caroline St. N., Waterloo, ON, N2L 2Y5, Canada}
\affiliation[b]{Department of Physics and Astronomy, University of Waterloo, Waterloo, ON, N2L 3G1, Canada}
\affiliation[c]{Department of Physics, Syracuse University, Syracuse, NY 13244-1130, U.S.A.}
\affiliation[d]{Raman Research Institute, C.V.Raman Avenue, Sadashivanagar, Bangalore 560 080, India}
\emailAdd{msaravani@perimeterinstitute.ca}
\emailAdd{saslanbeigi@perimeterinstitute.ca}
\emailAdd{rsorkin@perimeterinstitute.ca}

\abstract{
  We introduce a family of generalized d'Alembertian operators in
  $D$-dimensional Minkowski spacetimes $\Minkowski^D$ which are
  manifestly Lorentz-invariant, retarded, and non-local, the extent of
  the nonlocality being governed by a single parameter $\rho$.
  The prototypes of these operators arose in earlier work as averages of
  matrix operators meant to describe the propagation of a scalar field
  in a causal set.  We generalize the original definitions to produce an
  infinite family of ``Generalized Causet Box (GCB) operators''
  parametrized by certain coefficients $\{a,b_n\}$, and we derive the
  conditions on the latter needed for the usual d'Alembertian to be
  recovered in the infrared limit.  The continuum average of a GCB
  operator is an integral operator in $\Minkowski^D$, and it is these
  continuum operators that we mainly study.
  To that end, we compute their action on plane waves, or equivalently
  their Fourier transforms $g(p)$ [$p$ being the momentum-vector].
  For timelike $p$, $g(p)$ has an imaginary part whose sign
  depends on whether $p$ is past or future-directed.
  For small $p$, $g(p)$ is necessarily proportional to $p\cdot p$,
  but for large $p$ it becomes constant, raising the possibility of
  a genuinely Lorentzian perturbative regulator for quantum field
  theory in $\Minkowski^D$.
  We also address the question of whether or not the evolution defined by
  the GCB operators is stable, finding evidence that the original 4D
  causal set d'Alembertian is unstable, while its 2D counterpart is
  stable.}


\maketitle
\flushbottom
\section{Introduction}
%
%
%
Causal set theory postulates that the fundamental structure of
spacetime is that of a locally finite partially ordered set \cite{Sorkin_1}.
\footnote{
%
 Characterized mathematically, this is a set $C$ endowed with a
 binary relation $\prec$
 such that for all $x, y, z \in  C$ the following axioms are satisfied:
 (1) transitivity: $x\prec y$ $\&$ $y\prec z \Rightarrow x\prec z$;
 (2) irreflexivity: $x \nprec x$;
 (3):  local finiteness: $|\{y \in C|x \prec y \prec z\}| < \infty$.
 Thus a causal set (causet) is in a certain sense
 both
 Lorentzian [in virtue of (1) and (2)]
 and
 discrete [in virtue of (3)].
 }
%
Its
marriage of
discreteness
with
causal order
implies that physics cannot
remain local at all scales.
%
%
To appreciate why this should be,
let us
consider
how one might define a
notion of ``closeness''
in a causal set,
confining ourselves
to causal sets $C$ which are
obtained by randomly selecting points from a Lorentzian manifold $M$ and
endowing the selected points with
the causal relations inherited from the manifold.%
\footnote{
\label{poisson}
 This process is
 known as \emph{Poisson sprinkling}:
 Given a spacetime $M$, let the discrete subset of points, $C$, be one
 particular realization of a Poisson process in $M$, and let the
 elements of $C$ retain the causal relations they have when regarded as
 points of $M$.
 In order that the resulting precedence relation on $C$ approximately encode the
 metric of $M$, one must exclude spacetimes with closed causal curves,
 for example by requiring $M$ to be globally hyperbolic.
 }
Given such a causet, any intrinsically defined notion of closeness between
two elements of $C$ will reflect their Lorentzian distance
in the embedding spacetime.
%
But a small Lorentzian distance
between two
points of $M$
does \emph{not} mean that they
are confined to
a small neighbourhood
within $M$.
Rather, the second point can be ``arbitrarily distant'' from the first,
as long as it is located near to the lightcone of the latter.
Thus, an element of $C$ will inevitably possess \emph{very many}
``nearest neighbours'', no matter how that notion is formalized.
In this manner,
the concept of locality
provided by the topology of a continuous spacetime manifold is lost.

This
nonlocality
manifests itself concretely
when one seeks to describe
the wave propagation of a scalar field on a causal set
by defining
a discrete counterpart of the d'Alembertian operator, $\Box$.
For the aforementioned reasons,
it seems impossible to proceed in analogy with what one does when, for
example, one discretizes the Laplacian operator in a Riemannian spacetime.
%
%
Nevertheless,
a non-local
operator was
suggested
in \cite{Sorkin_2} which
on average
reproduces $\,\Box\,$ in the appropriate continuum
limit for $1+1$ dimensional Minkowski space $\Minkowski^2$
(i.e. for causets derived by sprinkling $\Minkowski^2$).
The expression introduced in \cite{Sorkin_2}
was generalized to $D=4$ dimensions in \cite{Benincasa_1}
and recently to arbitrary $D$ in \cite{Dowker_1}.

We shall denote a discrete causal set d'Alembertian
designed for $\Minkowski^D$
by $B_{\rho}^{(D)}$, where $\rho$
(dimensionally an inverse spacetime volume)
is a volume-scale
that
controls
the extent of the non-locality.
In the case of causal sets which are well-approximated by
$D$-dimensional Minkowski space
$\Minkowski^D$,
averaging $B_{\rho}^{(D)}$ over all
such causets
(i.e. averaging over all
sprinklings of $\Minkowski^D$ in the sense of footnote \ref{poisson})
leads to
a \emph{non-local}
and retarded
continuum operator $\Box^{(D)}_{\rho}$
defined in $\Minkowski^D$.
We shall refer to this operator as the \emph{continuum causal set d'Alembertian}.
Its crucial property
is that it reproduces the
usual d'Alembertian in the limit of zero non-locality scale:
 $\Box^{(D)}_{\rho}\phi\to\Box\phi$ as $\rho\to\infty$
for test-functions $\phi$ of compact support.

Although the causet operator $B^{(D)}_{\rho}$ is necessarily nonlocal,
one might expect that the range of its
nonlocality could be confined to the
discreteness scale itself.
In other words,
one might expect that $\rho\sim\ell^{-4}$,
$\ell$ being the
--- presumably Planckian ---
discreteness length.
However, one can also cite
reasons why
one might need to have
$\rho\ll\ell^{-4}$,
leading to a more long-range nonlocality.%
%
\footnote{
  The issue here concerns the behavior of $B^{(D)}_{\rho}$ for one
  particular sprinkling versus its behavior after averaging over all
  sprinklings.  The latter converges to $\Box$ as $\rho\to\infty$ but
  the former incurs fluctuations which grow larger as $\rho\to\infty$
  and which therefore will be sizable if $\rho$ is the sprinkling
  density, $\ell^{-4}$.  Which behavior is relevant physically?  In full
  quantum gravity some sort of sum over different causets will be
  involved, including in particular a sum over sprinklings.  Such a sum
  differs from a simple average and might or might not damp out the
  fluctuations, or they might cancel in other ways.  But if neither of
  these things happens, the only way out \cite{Sorkin_2} would be to
  choose $\rho$ small enough that the necessary averaging will occur
  within each individual causet.
}
Although these reasons are not conclusive, let us accept them provisionally.
A natural question then arises:
might such a ``mesoscopic'' nonlocality
show up at energy-scales accessible by current experiments?

Ideally, one would address this question in the fully discrete setting,
but it seems much easier to begin with the continuum version of the same question
by asking what changes when
the local operator $\Box$
is replaced by the nonlocal operator
$\Box^{(D)}_{\rho}$.
In this paper, we make a start on answering this question by analysing
the ``spectral properties'' (Fourier transform)
of a family of continuum operators $\Box^{(D)}_{\rho}$.
In Section \ref{org2D4D},
we discuss the continuum operators corresponding to the
original 2D \cite{Sorkin_2} and 4D \cite{Benincasa_1} causet
d'Alembertians,
and
in Section \ref{GenCB}
we
generalize the discussion
to an infinite
family of operators
parametrized by a set of coefficients,
$\{a,b_n\}$,
for which
we derive explicit equations
that ensure the usual flat space d'Alembertian is recovered in the
infrared limit.
Based on
 the UV behaviour of
these operators
(which we determine for all dimensions and coefficients $\{a,b_n\}$),
we
propose
a genuinely Lorentzian perturbative regulator for quantum field theory
(QFT).
Finally, we address the question of whether or not the evolution defined by
the (classical) equation
$\Box^{(D)}_{\rho}\phi=0$ is stable.
We devise a numerical method to test for stability and present strong evidence
that the original $4D$ causal set d'Alembertian is unstable in this
sense, while its 2D counterpart is stable.

Throughout the paper we use the metric signature ($-++\cdots$) and set $\hbar=c=1$.

\section{The Original 2D and 4D Causet d'Alembertians}
\label{org2D4D}
\begin{figure}
        \centering
        \includegraphics[width=0.6\hsize]{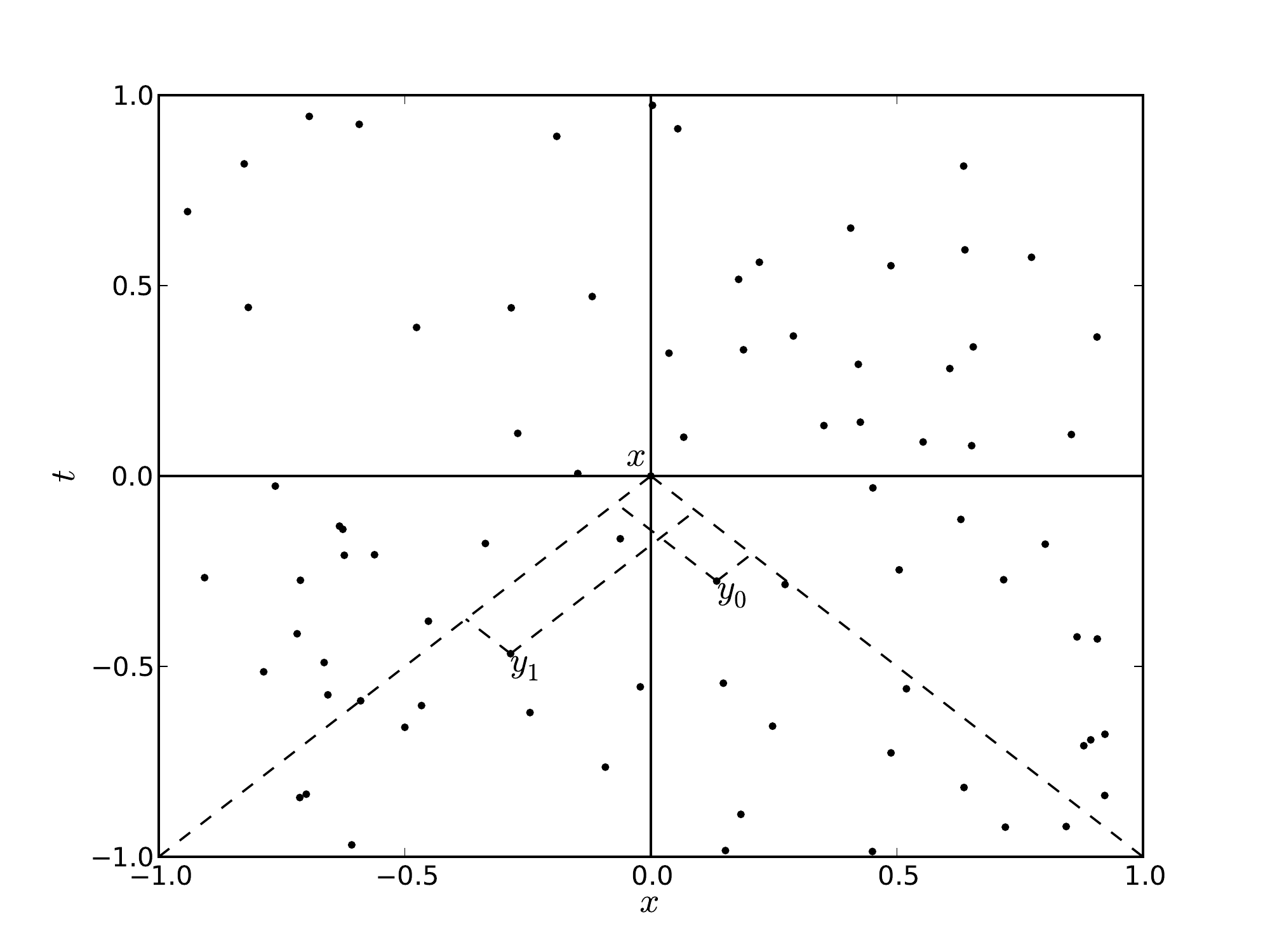}
        \caption{A Poisson sprinkling of $1+1$ Minkowski space at
          density $\rho=80$. Here $y_0$ is a $0$th neighbour of $x$
          because there are no elements which are both to the
          future of $y_0$ and the past of
          $x$.  Similarly, $y_1$ is a first neighbour of $x$. The
          contributions of the points $y_0$ and $y_1$ to
          $\rho^{-1}(B^{(2)}_{\rho}\Phi)(x)$ are $b^{(2)}_0\Phi(y_0)$
          and $b^{(2)}_1\Phi(y_1)$, respectively. The continuum limit,
          or rather average,
          of
          $(B_{\rho}^{(2)}\Phi)(x)$ can be understood as follows: fix
          the point $x$, keep sprinkling at density $\rho$ and compute
          $(B_{\rho}^{(2)}\Phi)(x)$ for every sprinkling. The average of
          all these values is equal to $(\Box_{\rho}^{(2)}\Phi)(x)$.}
        \label{2Dsprinkling}
\end{figure}

In this Section we discuss the  original
continuum causet d'Alembertians for
dimensions two \cite{Sorkin_2} and four \cite{Benincasa_1}.
Let us start by establishing some terminology.
Given any two elements $x,y$ of a causal set $C$,
we define
the {\it\/order interval\/}
Int$(x,y)$ between them as
the set of all elements
common to the (exclusive) future of $x$ and the (exclusive) past of $y$:
Int$(x,y)=\{z\in C|x\prec z \prec y\}$.
Notice that in our convention, Int$(x,y)$ does not include $x$ or $y$.
An element $y\prec x$ is then considered a past $n$th neighbour of $x$ if
Int$(y,x)$ contains $n$ elements.
For instance, $y$ is a $0$th neighbour of $x$ if Int$(y,x)$ is empty, a
first neighbour if Int$(y,x)$ contains one element, and so on (see Figure
\ref{2Dsprinkling} for an example).
We denote the set of all past $n$th neighbours of $x$ by $I_n(x)$.

Throughout the paper, we will only consider causal sets which are
obtained by Poisson sprinklings of Minkowski space at density $\rho$.
\subsection{2D}
\label{org2Dsec}
The original causet d'Alembertian
for dimension 2,
which we denote by $B^{(2)}_{\rho}$,
acts on a scalar field $\Phi(x)$ on the causal set in the following way
\cite{Sorkin_2}:
\beq
\rho^{-1}(B^{(2)}_{\rho}\Phi)(x)=a^{(2)}\Phi(x)+\sum_{n=0}^{2}b^{(2)}_n\sum_{y \in I_n(x)}\Phi(y),
\label{min2Ddis}
\eeq
where
\beq
  a^{(2)}=-2, \qquad
  b^{(2)}_0=4, \qquad
  b^{(2)}_1=-8, \qquad
  b^{(2)}_2=4.
  \label{coef2Dmin}
\eeq
Figure \ref{2Dsprinkling} illustrates
how $B^{(2)}_{\rho}$ is defined, given
a Poisson sprinkling of
2D Minkowski space $\Minkowski^2$.
The continuum
operator
$\Box_{\rho}^{(2)}$
is obtained by averaging $B^{(2)}_{\rho}$ over all
such
Poisson
sprinklings at density $\rho\,$:
\beq
 \rho^{-1}(\Box_{\rho}^{(2)}\Phi)(x)
  =
  a^{(2)}\Phi(x)+\rho\sum_{n=0}^{2}\frac{b^{(2)}_n}{n!}\int\limits_{J^{-}(x)}e^{-\rho V(x-y)}[\rho V(x-y)]^n\Phi(y)\,d^2y\;.
 \label{twopointthree}
\eeq
Here $J^{-}(x)$ denotes the causal past of $x$,
and $V(x-y)$ is the spacetime volume enclosed by
the past lightcone of $x$ and the future lightcone of $y$.
Note that $\Box_{\rho}^{(2)}$ is a \emph{retarded\/} operator,
in the sense that \eqref{twopointthree} uses information only from the causal past of $x$.
%
\begin{figure}
        \centering
        \begin{subfigure}[t]{0.45\hsize}
                \includegraphics[trim=50 60 50 40, clip=true, width=\hsize]{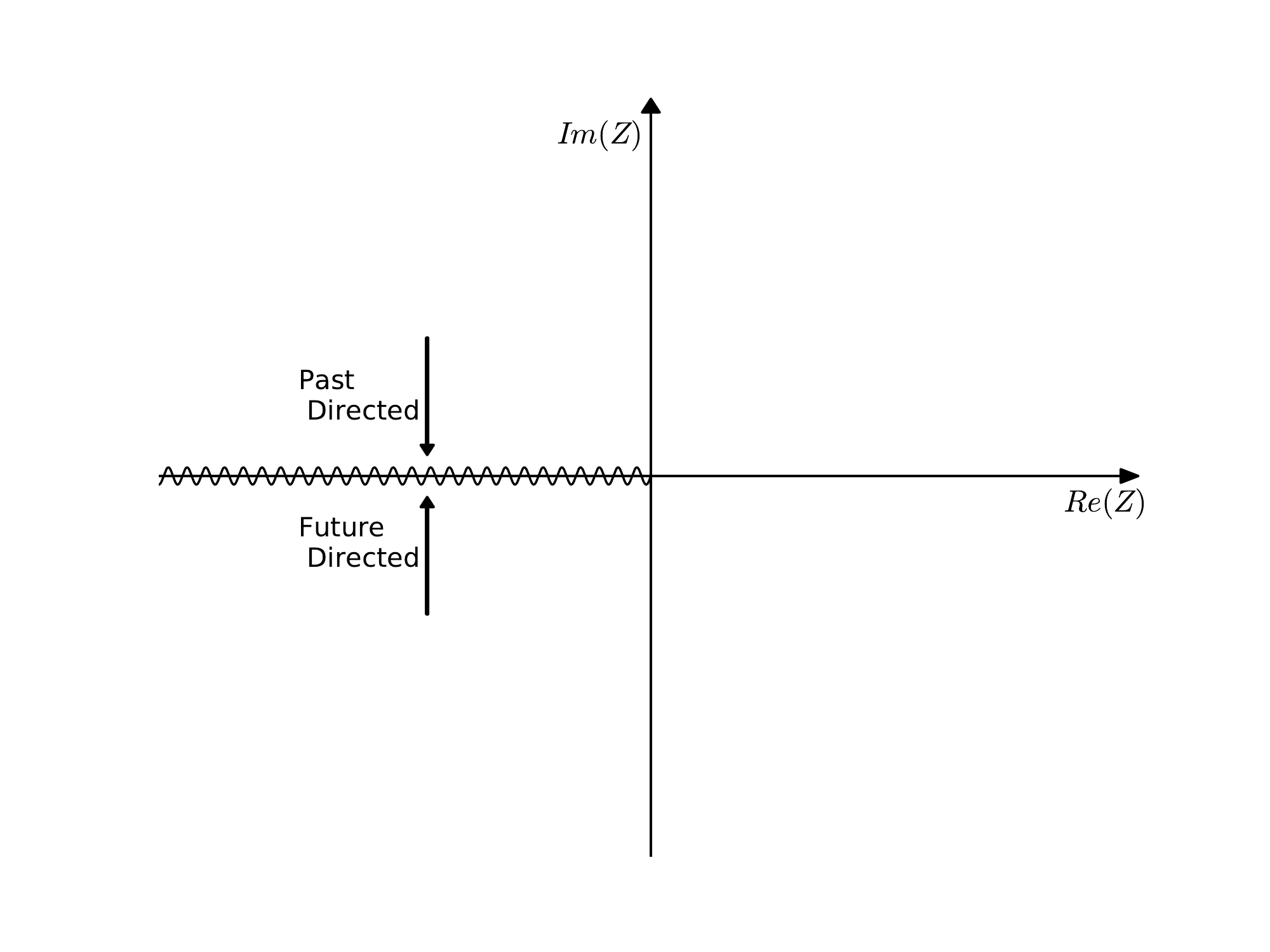}
                \caption{}
                \label{branchCut}
        \end{subfigure}%
        \begin{subfigure}[t]{0.60\hsize}
                \includegraphics[width=\hsize]{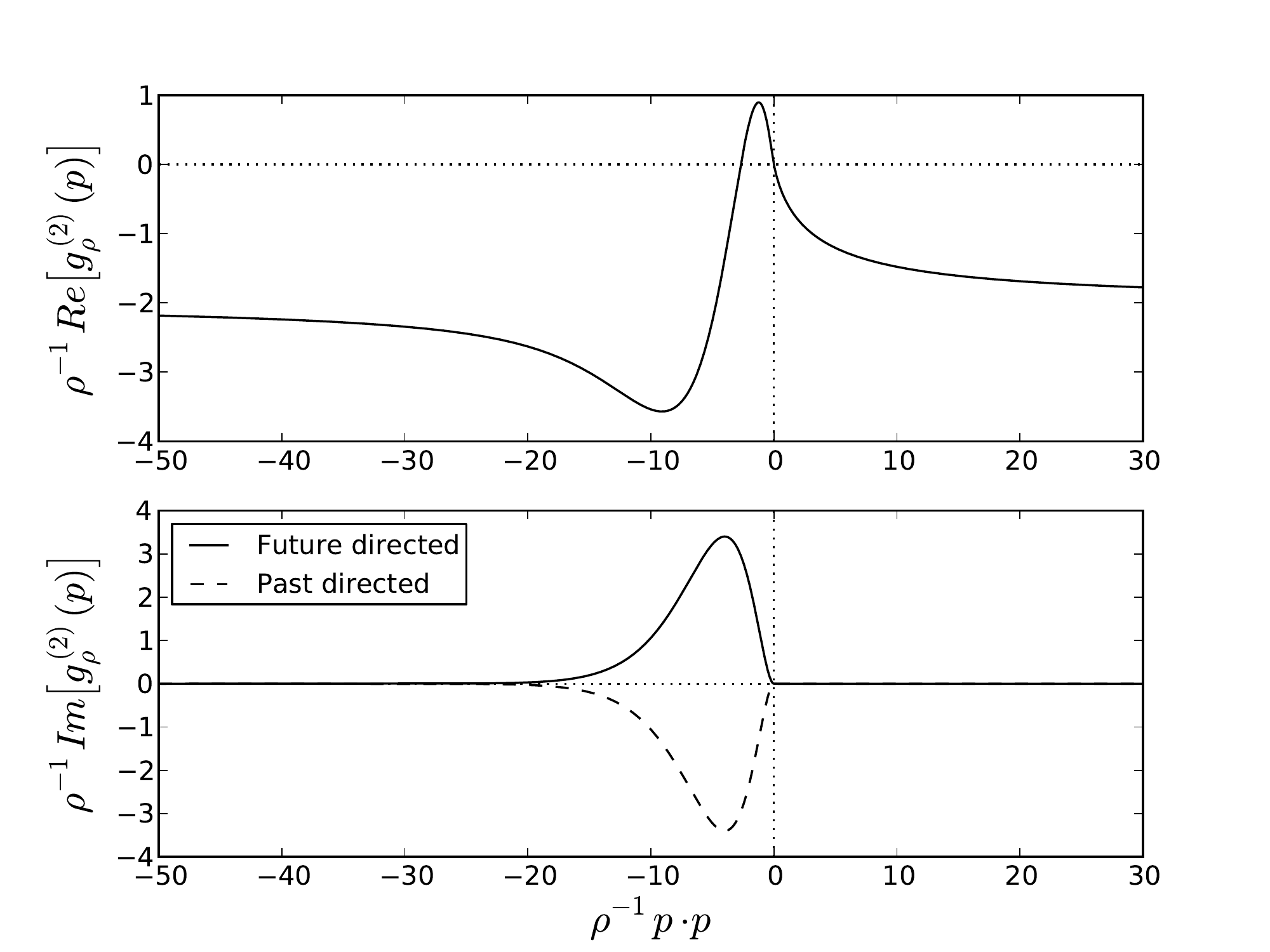}
                \caption{}
                \label{g2Dfig}
        \end{subfigure}
        \caption{(a) The principal branch of
          $\rho^{-1}g^{(2)}_{\rho}(p)$,
          which
          (for real p)
          depends only on
          $Z=\rho^{-1}\Lp\,$,
          and on sgn($p^0$) when $p$ is timelike.
          (b)
          The spectrum $g^{(2)}_{\rho}(p)$ of the original 2D
          continuum causet
          d'Alembertian for real momenta $p\,$.
          For spacelike momenta ($\Lp>0$),
          $g^{(2)}(p)$ is real.
          For timelike momenta, it
          is complex with an imaginary part
          whose sign is opposite
          for past-directed and future-directed momenta.}
        \label{fig:2D}
\end{figure}

The operator $\Box_{\rho}^{(2)}$ can be studied
by analysing its action on plane waves.  Due to translation symmetry of
Minkowski space,%
\footnote{
 This is why the volume $V$ in \eqref{twopointthree} is a function only
 of the difference, $x-y$.
}
any plane wave
$e^{ip\cdot x}$ is an eigenfunction of $\Box_{\rho}^{(2)}$
(provided that the integrals in \eqref{twopointthree} converge, so that
the left hand side is well defined):
\beq
 \Box_{\rho}^{(2)}e^{ip\cdot x} = g^{(2)}_{\rho}(p) e^{ip\cdot x},
\eeq
where $p\cdot x\equiv\eta_{\mu\nu}p^{\mu}x^{\nu}$ and
$\eta_{\mu\nu}=\text{diag}(-1,1)$.
%
%
Interestingly enough,
$g^{(2)}_{\rho}(p)$
in this case can be expressed in closed form:%
\footnote{
 This formula is derived in Appendix \ref{deriveExact2D}, using
 the general formalism developed in Section \ref{GenCB}.
}
\beq
  \rho^{-1}g^{(2)}_{\rho}(p)=-Ze^{Z/2}\text{E}_2(Z/2),
  \label{g2Dexact}
\eeq
where E$_2(z)$ is a generalized exponential integral function (see
e.g. 8.19 of \cite{DLMF}) and
\beq
  Z\equiv\rho^{-1}\Lp.
\eeq
Here,
as illustrated in Figure \ref{fig:2D},
E$_2(z)$ assumes its principal value,
with a branch cut along the negative real axis.
For real and spacelike momenta ($Z>0$), $g^{(2)}$ is real.
For real and timelike momenta ($Z<0$),
its value
above/below the branch cut corresponds to
past/future-directed momentum-vectors.
There, $g^{(2)}_{\rho}$ is complex
and changes to its complex conjugate across the cut.
That the spectrum is different for past and future-directed momenta
should come as no surprise, given that $\Box_{\rho}^{(2)}$ is retarded
by definition.
%
We will see in Section \ref{GenCB} that
these features
persist in all
dimensions and for a much broader class of causet d'Alembertians.

The infrared (IR) and ultraviolet (UV) behaviours
of $g^{(2)}_{\rho}(p)$
are easily deduced from the asymptotic
forms
of $E_2(Z)$
(see e.g. 8.11.2, 8.19.1, and 8.19.8 of \cite{DLMF}):
\begin{align}
  \rho^{-1}g^{(2)}_{\rho}(p)&\xrightarrow{Z\to0}-Z+\cdots\\
  \rho^{-1}g^{(2)}_{\rho}(p)&\xrightarrow{Z\to\infty}-2+\frac{8}{Z}+\cdots.
\end{align}
The first of these two equations
shows that the usual d'Alembertian $\Box$ is indeed
reproduced in the limit of zero non-locality.
The second equation,
on the other hand,
reveals a UV behaviour
quite unlike that of the usual d'Alembertian;
in Section \ref{IRGCB} it will lead us to propose a new regularization scheme
for quantum field theory.

An important question is whether the evolution defined by
$\Box_{\rho}^{(2)}\Phi=0$ is stable or not.  To a large extent
this is answered by the fact that the only zero
of  $g^{(2)}_{\rho}(p)$ occurs at
$Z=\rho^{-1}\Lp=0$.
%
To demonstrate this, we note that $g^{(2)}_{\rho}(p)$ has the
following representation (see e.g. 8.19.1  and 8.6.4 of \cite{DLMF}):
\beq
\rho^{-1}g^{(2)}_{\rho}(p)=-Zf(Z), \qquad
f(Z)\equiv\int_{0}^{\infty}\frac{te^{-t}}{t+Z/2}\ud t.
\eeq
It therefore suffices to prove that $f(Z)$ has no zeros when $Z\neq0$.
But the imaginary part of $f(Z)$ is
\beq
 \text{Im}(f(Z))
 =
 -\frac{\text{Im}(Z)}{2}\int_{0}^{\infty}\frac{te^{-t}}{\left[t+\frac{\text{Re}(Z)}{2}\right]^2+\left[\frac{\text{Im}(Z)}{2}\right]^2}\ud t.
 \label{eqImFz}
\eeq
Because the integral that multiplies $-\text{Im}(Z)/2$ in \eqref{eqImFz}
is strictly positive, $Zf(Z)$ could vanish only for real $Z$.
Obviously, it does vanish for $Z=0$, but elsewhere on the real axis, it remains nonzero,
as illustrated in Figure \ref{g2Dfig}.

What we have just proven is that a plane wave solves the equation
$\Box_{\rho}^{(2)}\Phi=0$ iff it solves the equation $\Box\Phi=0$.
To the extent that the general solutions of these two wave equations can
be composed of plane waves, they therefore share the same space of
solutions.  This, of course, is an important result in itself.  But it
also, a fortiori,  answers the stability question in the affirmative,
since we know that the evolution corresponding to $\Box$ is stable.

If there remains any doubt about stability or about the fact that both
$\Box\Phi=0$ and $\Box_{\rho}^{(2)}\Phi=0$ yield the same evolution, it
springs from a possible uncertainty about boundary conditions.  In the
usual situation (that of the ordinary d'Alembertian $\Box$), one understands
how to relate a general solution to its initial data on an arbitrary
Cauchy surface, and when $\Phi$ falls off suitably at infinity, its
total energy is defined and conserved.  From energy conservation,
stability also follows --- relative to the given choice of boundary
conditions.  On the other hand in the case of $\Box_{\rho}^{(2)}$, a
connection between solutions and Cauchy data remains to be found, as
does a better understanding of appropriate falloff conditions.
But absent some such boundary condition there is nothing to exclude complex
momenta $p$ that lead to exponential growth in time, e.g. an imaginary
multiple of a real lightlike vector.


For these reasons, we would like to discuss stability from a slightly
different angle, which also will be helpful when we come to deal with
the 4D case.
Quite generally, instabilities tend to be associated with exponentially
growing ``modes" (in this case plane waves).  Let us then {\it assume}
that we can take this as our criterion of (in)stability.  And to
exclude the kind of ``fake instability'' mentioned above, let us also
require any putative unstable mode, $\Phi(x)=e^{ip\cdot x}$, to be
bounded at spatial infinity in at least one Lorentz frame.
(Unfortunately we cannot say ``in all Lorentz frames'', since for a
plane wave, exponential growth in time induces exponential growth in
space via a Lorentz boost.)  We might hope that the condition just
formulated is equivalent to the following more natural one: consider
only solutions of $\Box_{\rho}^{(2)}\Phi(x)=0$ which have compact
support on every Cauchy hypersurface (compact spatial support in every
frame.)

Be that as it may, if this criterion is accepted, then we can establish
stability very simply in the present case, because an unstable mode,
$\Phi(x)=e^{ip\cdot x}$, is then precisely one such that $p$ possesses
a future-directed timelike imaginary part:
$p=p_R+ip_I$ with $p_I\cdot p_I<0$ and $p_I^0>0$.
This, however, is impossible for $Z=0$, as one sees from the equation
  $0=p\,\cdot\,p=p_R\cdot p_R-p_I\cdot p_I+2ip_R\cdot p_I$,
whose right-hand side has a strictly positive real part when $p_I$ is
timelike.
For logical completeness, we should also observe that \eqref{g2Dexact}
is valid for all complex $p$ whose imaginary parts are timelike and
future-directed.
(For more general complex momenta, the integral defining
$\Box_{\rho}^{(2)}\Phi$ might not converge, a circumstance that,
depending once again on the choice of falloff conditions, might or
might not impinge on the claimed identity between our solutions and
those of the ordinary wave equation.)


\subsection{4D}                 
\label{org4Dsec}
\begin{figure}
        \centering
        \begin{subfigure}[b]{0.45\hsize}
                \includegraphics[trim=30 10 30 30, clip=true, width=\hsize]{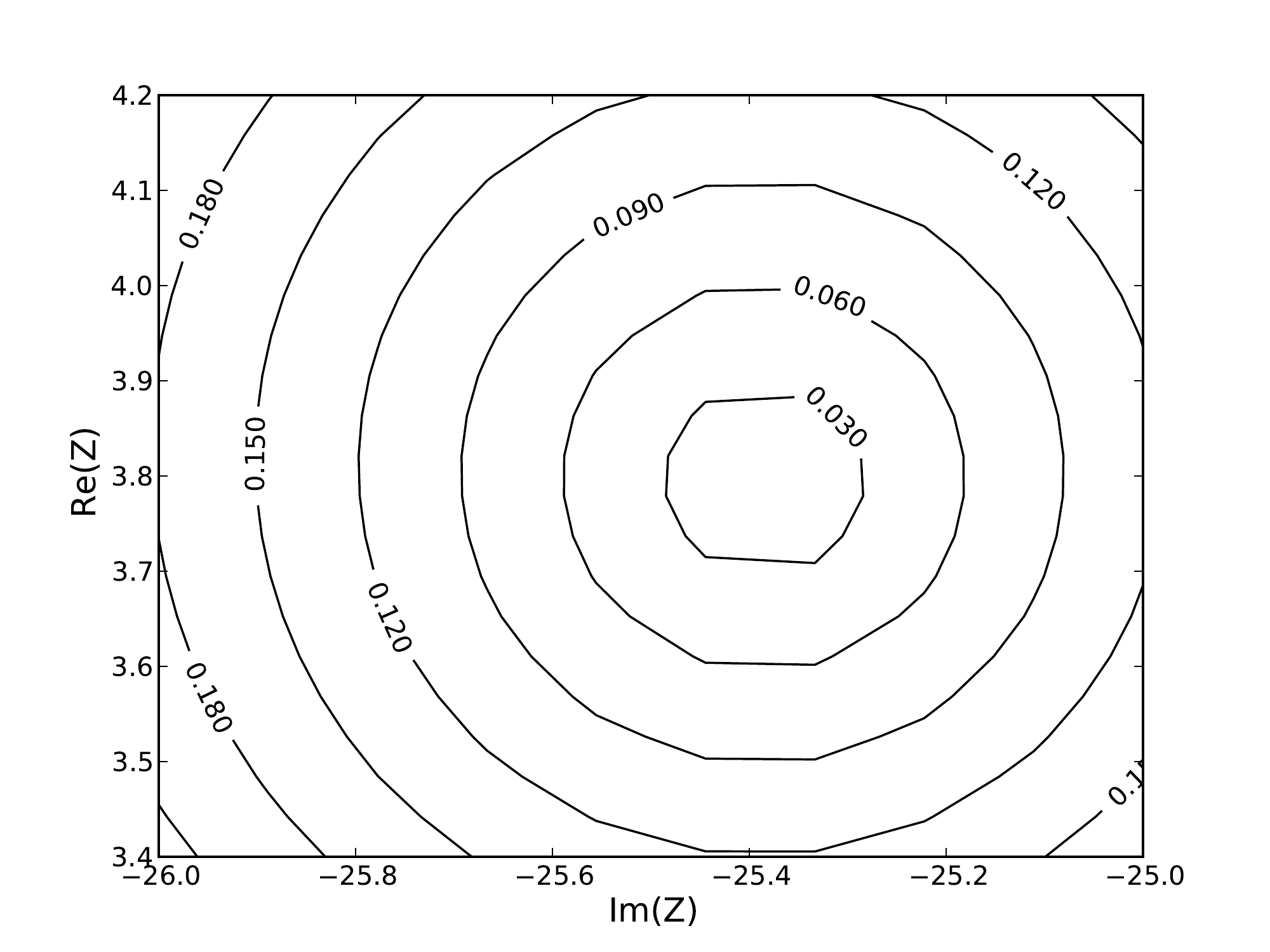}
                \caption{}
                \label{stability4D}
        \end{subfigure}%
        \begin{subfigure}[b]{0.57\hsize}
                \includegraphics[trim=0 0 30 30, clip=true, width=\hsize]{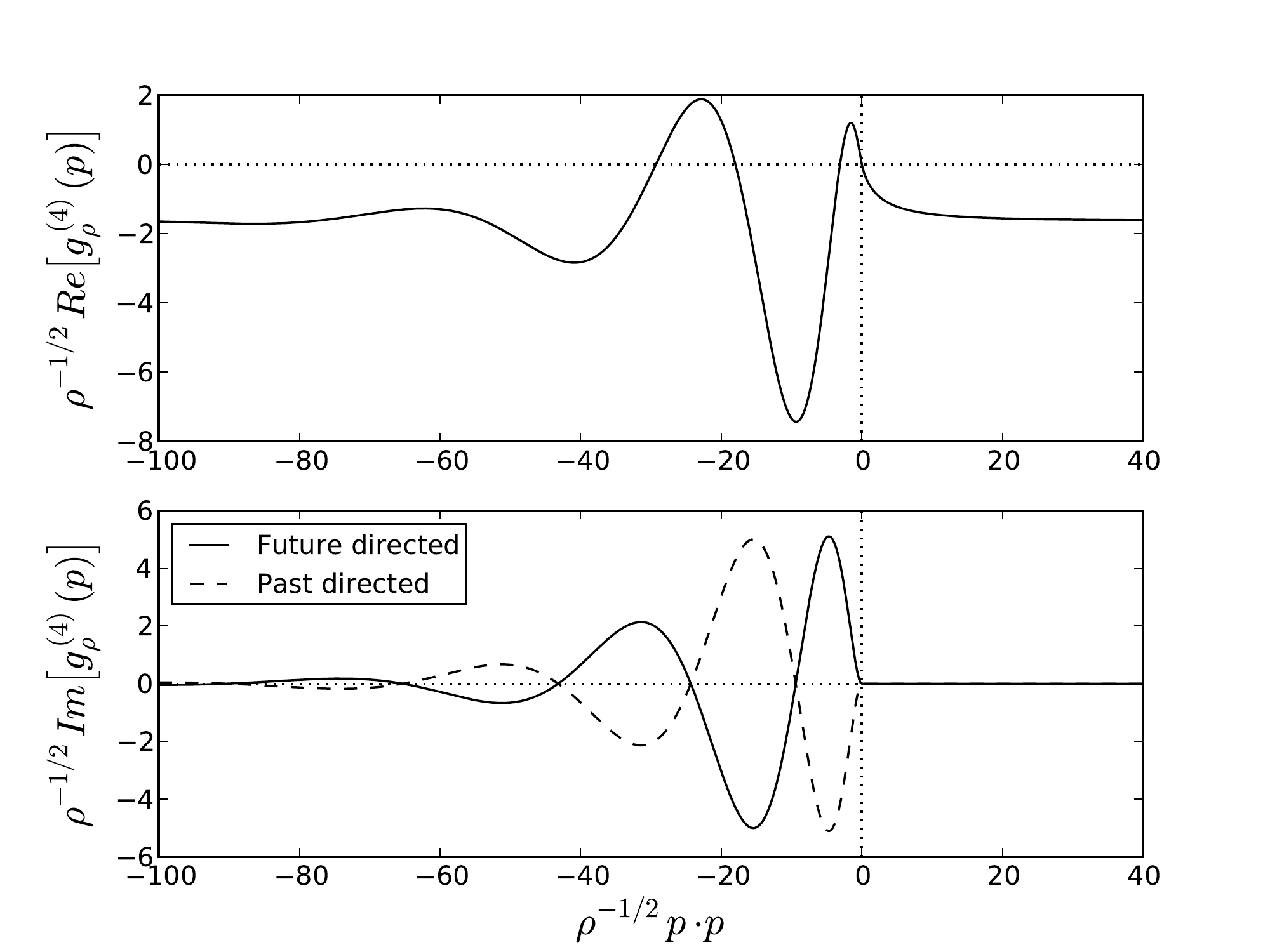}
                \caption{}
                \label{g4Dfig}
        \end{subfigure}
        \caption{(a) An unstable zero of $g^{(4)}_{\rho}(p)$.
          Contours of constant $|\rho^{-1/2}g^{(4)}_{\rho}|$ are plotted
          as a function of the real and imaginary parts of
          $Z=\rho^{-1/2}\Lp$.
          (b) Spectrum $g^{(4)}_{\rho}(p)$ of the
          original 4D causet d'Alembertian for real momenta $p$. For
          spacelike momenta ($\Lp>0$), $g^{(4)}(p)$ is real. For
          timelike momenta,
          it
          contains also an imaginary part
          whose sign is opposite
          for past-directed and future-directed
          momentum-vectors.}
        \label{fig:causetplot}
\end{figure}

The causet d'Alembertian
for dimension 4,
has the same general form as
that for $\Minkowski^2$,
but with different coefficients
\cite{Benincasa_1}~:
\beq
\rho^{-\frac{1}{2}}(B^{(4)}_{\rho}\Phi)(x)=a^{(4)}\Phi(x)+\sum_{n=0}^{3}b^{(4)}_n\sum_{y \in I_n(x)}\Phi(y),
\label{min4Ddis}
\eeq
where
\beq
a^{(4)}=-\frac{4}{\sqrt{6}}, \qquad
b^{(4)}_0=\frac{4}{\sqrt{6}}, \qquad
b^{(4)}_1=-\frac{36}{\sqrt{6}}, \qquad
b^{(4)}_2=\frac{64}{\sqrt{6}},\qquad
b^{(4)}_3=-\frac{32}{\sqrt{6}}.
\label{coef4Dmin}
\eeq
The
continuum average
$\Box_{\rho}^{(4)}$
then
also takes a similar form:
\beq
   \rho^{-\frac{1}{2}}(\Box_{\rho}^{(4)}\Phi)(x)
   =
   a^{(4)}\Phi(x)+\rho\sum_{n=0}^{3}\frac{b^{(4)}_n}{n!}\int\limits_{J^{-}(x)}e^{-\rho V(x-y)}[\rho V(x-y)]^n\Phi(y)d^4y.
\eeq
We will show in Section \ref{spectrumGen} that the ``spectrum'' of $\Box_{\rho}^{(4)}$,
as defined by
$\Box_{\rho}^{(4)}e^{ip\cdot x}=g^{(4)}_{\rho}(p)e^{ip\cdot x}$,
is given by
\beq
  \rho^{-1/2}g^{(4)}_{\rho}(p)=
  a^{(4)}+4\pi Z^{-1/2}\sum_{n=0}^{3}\frac{b^{(4)}_n}{n!}C_4^n
  \int_{0}^{\infty} s^{4n+2}e^{-C_4s^4}K_{1}(Z^{1/2}s)\ud s,
  \label{k14D}
\eeq
where $K_1$ is a modified Bessel function of the second kind and
\beq
  Z\equiv\rho^{-1/2}\Lp,\qquad
  C_4=\frac{\pi}{24}.
\eeq
All functions in \eqref{k14D} assume their principal values with branch cuts along the negative real axis.
Many properties of
the 2D function
$g^{(2)}_{\rho}(p)$
carry over to $g^{(4)}_{\rho}(p)\,$.
For timelike $p$, the value of $g^{(4)}_{\rho}(p)$ above/below the branch cut corresponds to
past/future-directed momenta,
and it changes to its complex conjugate across the cut.
Also, $g^{(4)}_{\rho}$ is real for spacelike momenta.
Figure \ref{g4Dfig} shows the behaviour of $g^{(4)}_{\rho}(p)$ for real momenta.

The IR and UV
behaviours
of $g^{(4)}_{\rho}(p)$,
which are derived in Sections \ref{IRGCB} and  \ref{UVGCB},
are given by
\begin{align}
  \rho^{-1/2}g^{(4)}_{\rho}(p)&\xrightarrow{Z\to0}-Z+\cdots\\
  \rho^{-1/2}g^{(4)}_{\rho}(p)&\xrightarrow{Z\to\infty}-\frac{4}{\sqrt{6}}+\frac{32\pi}{\sqrt{6}Z^2}+\cdots.
\end{align}
Again, the IR behaviour confirms that the usual d'Alembertian is reproduced in the limit of zero non-locality.
The UV limit has the form of a constant plus a term proportional to
$p^{-4}$.
The inverse of $g^{(4)}_{\rho}(p)$, which defines the retarded Green's function
in Fourier space, takes exactly the same form in the UV:
\beq
  \frac{\rho^{1/2}}{g^{(4)}_{\rho}(p)}\xrightarrow{Z\to\infty}-\frac{\sqrt{6}}{4}-\frac{2\pi\sqrt{6}}{Z^2}+\cdots.
\eeq
%
In any QFT based on $\Box_{\rho}^{(4)}$, the propagator associated with
internal lines in Feynman diagrams would
presumably
have the same UV
behaviour. Subtracting the constant term from the propagator (which
corresponds to subtracting a $\delta$-function in real space) would
then
render all loops finite.
This procedure could be
the basis of
a genuinely Lorentzian regularization and renormalization scheme for QFT.
We will discuss
these things more generally
in Sections \ref{UVGCB} and \ref{regGCB}.

We have only been able to address the question of stability by numerical
means in this case,
and we refer the reader to
Section \ref{stabGen}.
It turns out that $g^{(4)}_{\rho}(p)$ does in
fact have unstable
modes in the sense that
there exist
complex
momentum-vectors $p$
which satisfy $g^{(4)}_{\rho}(p)=0$,
and whose
imaginary parts
are timelike and future-directed.
Such a mode corresponds to a complex zero of $g^{(4)}_{\rho}$ in the
complex $Z$-plane, and
Figure \ref{stability4D} shows one such zero (the other one being its
complex conjugate).
%

\section{The Generalized Causet Box (GCB) Operators}
\label{GenCB}
The key property of the causet d'Alembertians introduced in the previous
Section is that they reproduce $\Box$ in the continuum-averaged (averaged over
all sprinklings) and local ($\rho\to\infty$) limit.
In this Section, we explore
a larger family of
operators $B^{(D)}_{\rho}$ which share the same property.
We
place
the
following conditions on $B^{(D)}_{\rho}$:
\begin{enumerate}
  \item \textbf{Linearity}: when $B^{(D)}_{\rho}$ acts on a scalar field
    $\Phi$, the result at an element $x$ of the causet should be a
    linear combination of the values of $\Phi$ at other elements $y$
    (possibly including $x$ itself).  This is a natural requirement
    because $\Box$ itself is linear.

  \item \textbf{Retardedness}: $(B^{(D)}_{\rho}\Phi)(x)$ should depend
    only on $\Phi(y)$, with $y$ in the causal past of $x$.  This
    requirement allows for a consistent evolution of
    a partial solution specified on any ``downward closed'' subset of
    the causet.

\item \textbf{Label invariance}:  $B^{(D)}_{\rho}$ should be invariant
  under relabellings of causal set elements.  This is the discrete
  analogue of  general covariance.
\item \textbf{Neighbourly democracy}: all $n$th neighbours of $x$ should
  contribute to $(B^{(D)}_{\rho}\Phi)(x)$ with the same coupling.
\end{enumerate}
Considering all these requirements, $(B^{(D)}\Phi)(x)$ can be expressed
in the following general form
\beq
  \rho^{-\frac{2}{D}}(B^{(D)}_{\rho}\Phi)(x)=a\Phi(x)+\sum_{n=0}^{\Lmax}b_n\sum_{y \in I_n(x)}\Phi(y),
  \label{genDisBox}
\eeq
where $\{a,b_n\}$ are dimensionless coefficients and $I_n(x)$ is the set
of all $n$th neighbours to the past of $x$ (see beginning of Section
\ref{org2D4D}).
This is a straightforward generalization of \eqref{min2Ddis} and
\eqref{min4Ddis}, where we have now allowed ourselves up to $\Lmax$
neighbours.
We will soon see that recovering $\Box$ requires keeping a
\emph{minimum} number of layers: e.g. $\Lmax\ge2$ in 2D and $\Lmax\ge3$
in 4D.  The original 2D and 4D proposals are then the minimal cases in
this sense.

The continuum-average $\Box^{(D)}_{\rho}$ of $B^{(D)}_{\rho}$ acts on a
scalar field $\Phi(x)$ in the following way:
\beq
  \rho^{-2/D}(\Box^{(D)}_{\rho}\Phi)(x)
   =
   a\Phi(x)+\rho\sum_{n=0}^{\Lmax}\frac{b_n}{n!}\int\limits_{J^{-}(x)}e^{-\rho V(x,y)}[\rho V(x,y)]^n\Phi(y) d^Dy.
  \label{tatu2}
\eeq
Here as before, $J^{-}(x)$ denotes the causal past of $x$,
while $V(x,y)$ is the spacetime volume enclosed by the
past light cone of $x$ and the future light cone of $y$.

The occurrence of the factor $e^{-\rho V}$ in \eqref{tatu2} shows that
the parameter $\rho$ (which dimensionally is an energy-density)
functions as a kind of ``nonlocality scale'' controlling the distance
over which the operator $\Box^{(D)}_{\rho}$ acts.  As our definitions
stand so far, this nonlocality-scale directly reflects the fundamental
discreteness-scale,
because \eqref{tatu2} was derived under the
assumption that $\rho$ was the sprinkling-density in $\Minkowski^D$.
However it turns out that one can decouple the two scales by tweaking
the definition \eqref{genDisBox} in such a way as to produce a more
general causet operator
whose
sprinkling-average reproduces the same continuum operator \eqref{tatu2},
even when
$\rho$ is smaller than the sprinkling density.
With this operator,
the nonlocality can
extend over a much greater distance than that of the fundamental
discreteness-scale.  Although modifying $B^{(D)}_{\rho}$
in this way has the disadvantge of introducing a
second, independent length scale, it allows one to overcome a potential
difficulty pointed out in \cite{Sorkin_2}, namely that \eqref{genDisBox}
with fixed coefficients
leads to fluctuations in $(B^{(D)}_{\rho}\Phi)(x)$ which grow with
$\rho$, rather than diminishing.  We have provided the definition of
this ``tweaked'' operator
and the derivation of its continuum
average in Appendix~\ref{dampFluct};
but henceforth,
we will concern
ourselves exclusively with the continuum operator $\Box^{(D)}_{\rho}$,
without worrying about its relationship with any underlying
discreteness.  Correspondingly,
\emph{$\rho$ will henceforth denote a non-locality-scale with no necessary relation to any discreteness scale}.

\subsection{Spectrum}
\label{spectrumGen}
That any plane wave $e^{ip\cdot x}$ is an eigenfunction of
$\Box^{(D)}_{\rho}$ in $\Minkowski^D$ follows from
translational symmetry: $V(x,y)=V(x-y)$.
It can be shown in fact that
\begin{align}
 \Box^{(D)}_{\rho} \; e^{ip\cdot x}&=g^{(D)}_{\rho}(p) \; e^{ip\cdot x},\\
 \rho^{-2/D}g^{(D)}_{\rho}(p)&=a+\sum_{n=0}^{\Lmax}\frac{(-1)^n\rho^{n+1}}{n!}
 \, b_n\, \frac{\partial^n}{\partial\rho^n}\chi(p,\rho),\label{gGen}\\
  \chi(p,\rho) &= \int\limits_{J^{+}(0)} e^{-\rho V(y)} \; e^{-ip\cdot y} \; d^Dy,
\end{align}
where $V(y)=V(O,y)$ is the spacetime volume enclosed by
the past light cone of $y$ and the future light cone of the origin:
\beq
  V(y) = C_D \;  |y\cdot y|^{D/2}, \qquad
  C_D=\frac{\left(\frac{\pi}{4}\right)^{\frac{D-1}{2}}}{D\Gamma(\frac{D+1}{2})}\ .
\eeq

Evaluating $\chi(p,\rho)$ amounts to computing the Laplace transform of
a retarded, Lorentz-invariant function, which has been done
in \cite{LTRLI}. It follows from their result that
\beq
  \chi(p,\rho)=2(2\pi)^{D/2-1}(\Lp)^{\frac{2-D}{4}}\int_{0}^{\infty}s^{D/2}e^{-\rho C_Ds^D}K_{\frac{D}{2}-1}(\sqrt{\Lp} s)\ud s,
  \label{chiP}
\eeq
where $K_{\nu}$ is the modified Bessel function of the second kind.
All functions in \eqref{chiP} assume their principal values, with a branch cut along the negative real axis.
This result is valid for all $p$ whose imaginary part is timelike and future-directed,
i.e.
$p_I\cdot p_I<0$ and $p_I^0>0$,
where
$p=p_R+ip_I$
and the
Lorentzian norm is given by $p\cdot p=p_R\cdot p_R-p_I\cdot p_I+2ip_R\cdot p_I$.
For momenta satisfying these conditions,
the integral that defines $\chi(p,\rho)$,
and consequently $\Box^{(D)}_{\rho}e^{ip\cdot x}$, is absolutely
convergent.
Plugging \eqref{chiP}
into
\eqref{gGen} we find
\beq
  \boxed{\rho^{-2/D}g^{(D)}_{\rho}(p)=a+2(2\pi)^{D/2-1}Z^{\frac{2-D}{4}}\sum_{n=0}^{\Lmax}\frac{b_n}{n!}C_D^n
  \int_{0}^{\infty} s^{D(n+1/2)}e^{-C_Ds^D}K_{\frac{D}{2}-1}(Z^{1/2}s)\ud s,}
  \label{geng}
\eeq
where $Z$ is a dimensionless quantity defined by
\beq
   Z\equiv\rho^{-\frac{2}{D}}\Lp.
\eeq

For real $p=p_R$, $g^{(D)}_{\rho}(p)$ can be defined by first adding a small
future-pointing
and timelike imaginary part $p_I^{\epsilon}$ to $p_R$,
and then taking the limit as $p_I^{\epsilon}$ shrinks:
\beq
   g^{(D)}_{\rho}(p_R):=\lim_{\epsilon\to0^{+}}g^{(D)}_{\rho}(p_R+ip_I^{\epsilon}), \qquad
   p_I^{\epsilon}\cdot p_I^{\epsilon}=-\epsilon^2.
\eeq
When $p_R$ is timelike, this amounts to changing $Z=\rho^{-\frac{2}{D}}p_R\cdot p_R$ on the right hand side of \eqref{geng} to
$Z+ i\epsilon$ for past-directed, and $Z- i\epsilon$ for future-directed $p_R$. This is illustrated in Figure \ref{branchCut}.
Because of the appearance of $Z^{1/2}$ in \eqref{geng} and the fact that
$K_{\nu}(\bar{z})=\overline{K_{\nu}(z)}$, it follows for timelike $p$
that
\beq
   g^{(D)}_{\rho}(-p)=\overline{g^{(D)}_{\rho}(p)}.
\label{cmplxConj}
\eeq
Therefore,
\emph{$g^{(D)}_{\rho}(p)$ differs for past- and future-directed timelike $p$}.
This is to be expected,
since requiring $\Box^{(D)}_{\rho}$ to be retarded builds in a direction of time.
For spacelike momenta ($Z>0$), $g^{(D)}_{\rho}(p)$ is real, as follows
from the fact that $K_{\nu}(z)$ is real when $\nu$ is real and
$\text{ph}(z)=0$ \cite{DLMF}.

\subsection{IR Behaviour}
\label{IRGCB}
We want to choose
the coefficients $a$ and $b_n$ so that
the usual d'Alembertian operator is recovered in the limit of zero
non-locality:
\beq
  \lim_{\rho\to\infty}\Box^{(D)}_{\rho}\phi=\Box\phi.
  \label{IR_1}
\eeq
This requirement is equivalent to demanding
\beq
  g^{(D)}_{\rho}(p) \xrightarrow{Z\to 0}
  -\Lp  \ .
  \label{IR_2}
\eeq
In Appendix \ref{IRdet},
we derive
equations for $a$ and $b_n$ which guarantee this behaviour
for an arbitrary spacetime dimension D.
We expand $Z^{\frac{2-D}{4}}K_{\frac{D}{2}-1}(Z^{1/2}s)$
on the right hand side of \eqref{geng} about $Z=0$,
and arrange $a, b_n$ so that the terms
which grow faster than $Z$ vanish,
while the coefficient of the term proportional to $Z$ is $-1$.
We state the main results here and refer the reader to Appendix
\ref{IRdet} for the details.

In \textbf{even dimensions}, letting $D=2N+2$ with $N=0,1,2,\dots$,
the equations that need to be satisfied are
\begin{subequations}
        \begin{align}
                \sum_{n=0}^{\Lmax}\frac{b_n}{n!}\Gamma(n+\frac{k+1}{N+1})&=0, \qquad k=0,1,\dots,N+1\label{even1}\\
                a+\frac{2(-1)^{N+1}\pi^N}{N!D^2C_D}\sum_{n=0}^{\Lmax}b_n\psi(n+1)&=0\label{even2},\\
                \sum_{n=0}^{\Lmax}\frac{b_n}{n!}\Gamma(n+\frac{N+2}{N+1})\psi(n+\frac{N+2}{N+1})&=\frac{2(-1)^N(N+1)!}{\pi^N}D^2C_D^{\frac{N+2}{N+1}},\label{even3}
        \end{align}
\end{subequations}
where $\psi(n)$ is the digamma function.
Equations \eqref{even1} and \eqref{even3} determine $b_n$,
after which \eqref{even2} fixes $a$.
The minimum number of terms required to solve these equations
is determined by $\Lmax\ge N+2$.
In 2D and 4D
in particular,
keeping this minimum number of terms leads to the solutions
\eqref{coef2Dmin} and
\eqref{coef4Dmin}, respectively.

In \textbf{odd dimensions}, letting $D=2N+1$ with $N=0,1,2,\dots$,
the equation are
\begin{subequations}
        \begin{align}
                \sum_{n=0}^{\Lmax}\frac{b_n}{n!}\Gamma(n+\frac{2k+2}{2N+1})&=0, \qquad k=0,1,\dots,N\label{odd1}\\
                a+\frac{(-1)^{N}\pi^{N+\frac{1}{2}}}{DC_D\Gamma(N+\frac{1}{2})}\sum_{n=0}^{\Lmax}b_n&=0\label{odd2},\\
                \sum_{n=0}^{\Lmax}\frac{b_n}{n!}\Gamma(n+\frac{2N+3}{2N+1})&=\frac{4(-1)^{N-1}\Gamma(N+\frac{3}{2})}{\pi^{N+\frac{1}{2}}}DC_D^{\frac{2N+3}{2N+1}}.\label{odd3}
        \end{align}
\end{subequations}
Similarly to the even case,
Equations \eqref{odd1} and \eqref{odd3} determine $b_n$, after which
\eqref{odd2} fixes  $a$.
The minimum number of terms is determined by $\Lmax\ge N+1$.
\subsection{UV Behaviour and the Retarded Green's Function}
\label{UVGCB}
The UV behaviour of $g^{(D)}_{\rho}(p)$, as derived in Appendix \ref{UVdet}, is
\beq
  \rho^{-2/D} g^{(D)}_{\rho}(p)
    \xrightarrow{Z\to\infty}
    a + 2^{D-1}\pi^{\frac{D}{2}-1}\Gamma(D/2)\, b_0\, Z^{-\frac{D}{2}}+\cdots.
  \label{uvGen}
\eeq
Thus, $g_{\rho}^{(D)}(p)$ behaves as a constant plus a term
proportional to $(\Lp)^{-D/2}$.  Let us explore the consequences
of this fact for the retarded Green's function $G_R(x,y)$ associated
with $\Box_{\rho}^{(D)}$, which satisfies the usual equation
\beq
   \Box_{\rho}^{(D)}G_R(x,y)=\delta^{(D)}(x-y),
\eeq
subject to the boundary condition $G_R(x,y)=0$ $\forall$ $x\nsucceq y$.

\begin{figure}
        \centering
        \begin{subfigure}[t]{0.5\hsize}
                \includegraphics[trim=50 60 50 40, clip=true, width=\hsize]{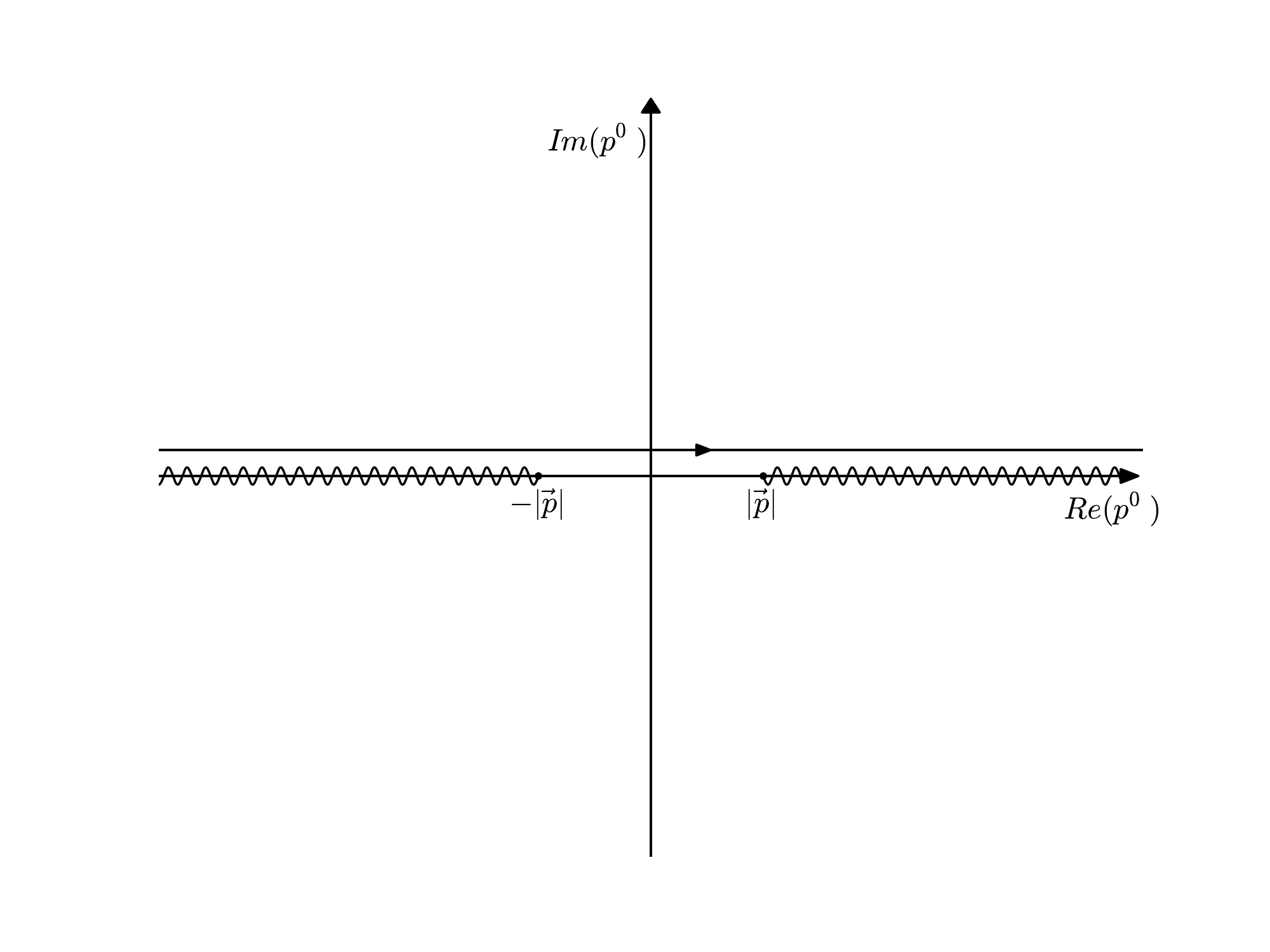}
                \caption{}
                \label{greenIntg}
        \end{subfigure}%
        \begin{subfigure}[t]{0.50\hsize}
                \includegraphics[trim=50 40 50 40, clip=true, width=\hsize]{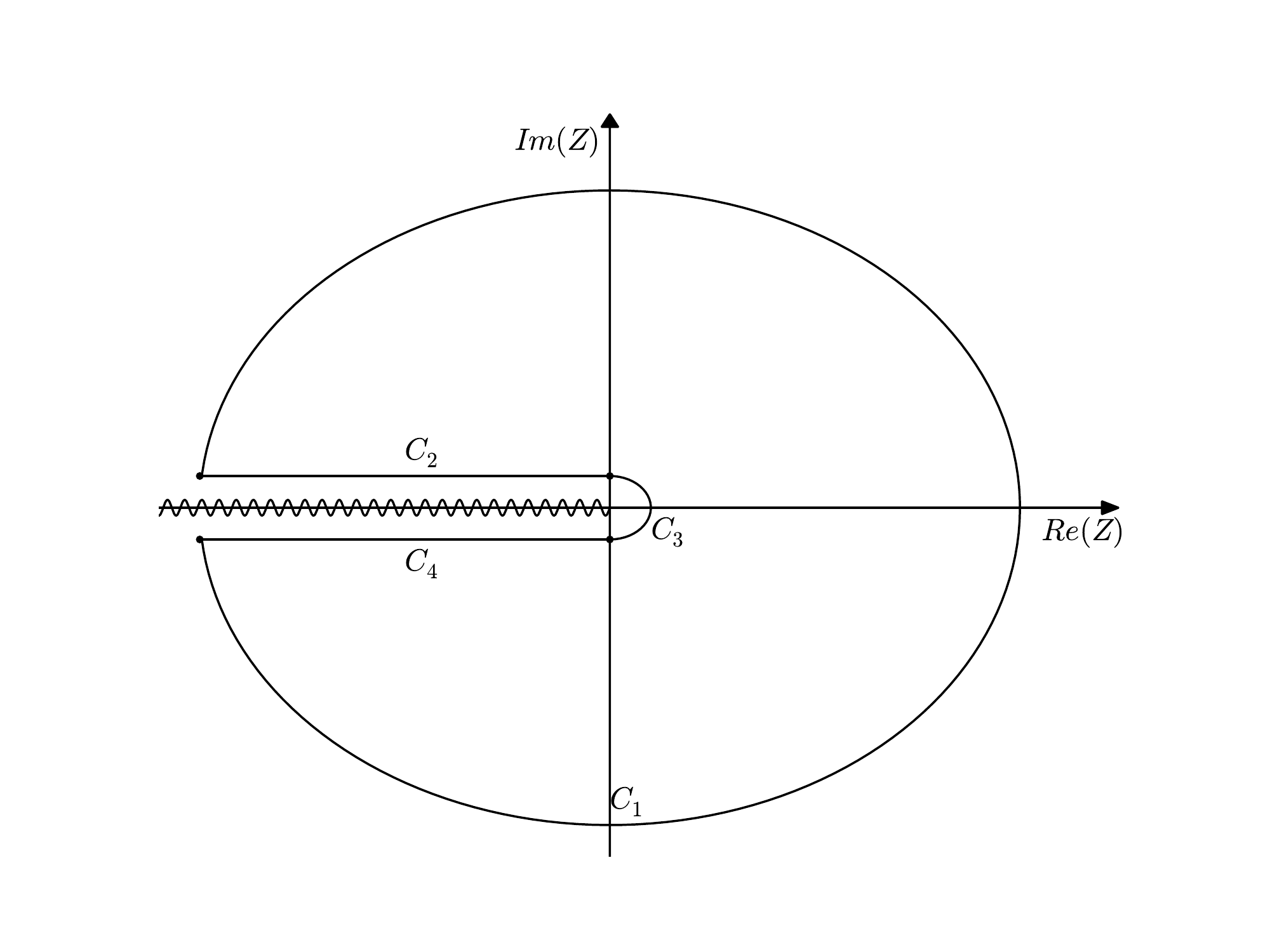}
                \caption{}
                \label{stabilityCont}
        \end{subfigure}
        \caption{(a) The integration path in the complex $p^0$ plane
          which defines the retarded Green's function. (b) The contour of
          integration used for counting the unstable modes of
          $\Box^{(D)}_{\rho}$.  The direction of integration is taken
          to be counter-clockwise.}
        \label{fig:contour}
\end{figure}

Of course, translation invariance implies $G_R(x,y)=G_R(x-y)$.
%
The Fourier transform $\tilde{G}_R(p)$ of $G_R(x-y)$ is given by the
reciprocal
of $g_{\rho}^{(D)}(p)$:
\beq
   G_R(x-y)=\int \frac{d^Dp}{(2\pi)^D}\tilde{G}_R(p) e^{ip\cdot (x-y)=}\int \frac{d^Dp}{(2\pi)^D}\frac{1}{g_{\rho}^{(D)}(p)}e^{ip\cdot (x-y)}.
\label{grnFourier}
\eeq
Figure \ref{greenIntg} shows the path of integration in the complex
$p^0$ plane.
When $g_{\rho}^{(D)}(p)$ has no zero in complex plane apart
from at $\Lp=0$, this
choice of contour ensures that $G_R$ is indeed retarded.
As we will
argue in the next section, the presence of such zeros
implies
that
evolution defined by $\Box_{\rho}^{(D)}$ is unstable.
Therefore, we shall ignore these cases for our current discussion.

The behaviour of $G_R(x-y)$ in the coincidence limit $x\to y$ is
determined by the behaviour of $\tilde{G}_R(p)$ at large momenta:
\beq\label{meh1}
  \rho^{2/D}\tilde G_R(p)
   \xrightarrow{Z\to\infty} \;
    \frac{1}{a} \; - \; 2^{D-1}\pi^{\frac{D}{2}-1}\Gamma(D/2)\; \frac{b_0}{a^2} \;Z^{-\frac{D}{2}}+\cdots
\eeq
Here we have assumed $a\neq0$.
When $a=0$, $\tilde{G}_R(p)$ scales as $p^{D}$ for large momenta, a
badly divergent UV behaviour.
Therefore we will confine ourselves to cases where $a\neq0$.

The constant term $\frac{1}{a}$ represents a $\delta$-function in real
space.  The other terms in the series have the form
$\int d^Dp~p^{-nD}, n=1,2,\cdots$,
and it can be shown that they are all finite.
It then
looks like subtracting $\frac{1}{a} \delta^{(D)}(x-y)$ from
$\rho^{2/D}G_R(x-y)$ must result in a completely smooth function in the
coincidence limit, and  we will now show this is indeed the case.

Although $D=4$ is the dimension of greatest interest, the proof which we
shall present is valid in all even dimensions.
Let us define
\beq
   \rho^{2/D}G(x-y)\equiv \rho^{2/D}G_R(x-y)-\frac{1}{a}\delta^{(D)}(x-y).
\eeq
Our task is then to show $G(x-y)$ is a smooth function at
$x=y$.
It follows from \eqref{grnFourier} that
\beq\label{meh2}
\rho^{2/D} G(x-y)=\int \frac{d^Dp}{(2\pi)^D}\left[\frac{1}{\rho^{-2/D}g_{\rho}^{(D)}(p)}-\frac{1}{a}\right]e^{ip\cdot (x-y)}.
\eeq
Because $G_R(x-y)$ is retarded by definition,
\beq\label{meh3}
   \int \frac{d^Dp}{(2\pi)^D}\frac{1}{g_{\rho}^{(D)}(p)}e^{ip\cdot(x-y)}=0, \qquad x\nsucceq y \ .
\eeq
From this it follows for all $x\succ y$ that
\beq\label{meh5}
  \int
  \frac{d^Dp}{(2\pi)^D}\frac{1}{\overline{g_{\rho}^{(D)}(p)}}e^{ip\cdot(x-y)}=\int
  \frac{d^Dp}{(2\pi)^D}\frac{1}{g_{\rho}^{(D)}(p)}e^{-ip\cdot(x-y)}\overset{x\succ
    y}{=}0 \ ,
\eeq
where the first equality is obtained by changing $p\rightarrow -p$ and
then using \eqref{cmplxConj}, and the second equality is a direct consequence
of \eqref{meh3}
with $x$ and $y$ interchanged.
Returning to \eqref{meh2}, and subtracting zero in the form of \eqref{meh5},
we obtain
\begin{align}
  G(x-y)&\overset{x\succ y}{=}\int\frac{d^Dp}{(2\pi)^D}\left[\frac{1}{g_{\rho}^{(D)}(p)}-\frac{1}{\overline{g_{\rho}^{(D)}(p)}}\right]e^{ip\cdot(x-y)}\\
  &=\int_{p^2<0}\frac{d^Dp}{(2\pi)^D}\left[\frac{1}{g_{\rho}^{(D)}(p)}-\frac{1}{\overline{g_{\rho}^{(D)}(p)}}\right]e^{ip\cdot(x-y)}
  \ ,
\label{meh6}
\end{align}
where the second equality is true because $g_{\rho}^{(D)}(p)$ is real for space-like momenta.
(Note that the $\frac{1}{a}$ term contributes only when $x=y$.)
In what follows, we let
\beq
  \rho^{-2/D}g^{(D)}_{\rho}(p)\equiv\tilde{g}(Z),
  \label{tildeg}
\eeq
as given in the right hand side of \eqref{geng}.

The integral in \eqref{meh6} can be divided into two integrals over
$p^0>0$ and $p^0<0$. For a fixed sign of $p^0$, $g_{\rho}^{(D)}(p)$ is
only a function of $p\cdot p$, making \eqref{meh6} the Laplace transform
of a Lorentz-invariant function.
Similarly to how we derived \eqref{chiP},
we
use the result of \cite{LTRLI} to compute $G(x-y)$:
\begin{align}
  \rho^{2/D}G(x-y)\overset{x\succ y}{=}&\frac{2}{\pi(2\pi)^{D/2}}\int_0^{\infty}d\xi~\xi^{D/2}\notag\\
  &\times\text{Re}\left[\left(\sqrt{s_{\epsilon}^2}\right)^{1-\frac{D}{2}}
    K_{\frac{D}{2}-1}(\sqrt{s_{\epsilon}^2}\,\xi)
   \left(\frac{1}{\tilde g(-\xi^2+i\epsilon)}-\frac{1}{\overline{\tilde g(-\xi^2+i\epsilon)}}\right)\right],
   \label{meh7}
\end{align}
where $s_{\epsilon}^2=-(t_x-t_y+i\epsilon)^2+|\vec r_x-\vec r_y|^2$ and
$\epsilon$ is a small positive number which should be taken to zero at
the end of calculations.
When $x-y$ is timelike and future-directed, we can let $\sqrt{s_{\epsilon}^2}=-i\tau_{xy}$ where $\tau_{xy}>0$.
Using properties of Bessel functions (see e.g. 10.27.9 of \cite{DLMF}),
\eqref{meh7} can be simplified into the following form for even D:
\begin{align}
   \rho^{2/D}G(x-y)
   &\overset{x\succ y}{=}
   \frac{-i(-1)^{\frac{D}{2}}\tau_{xy}^{1-\frac{D}{2}}}{(2\pi)^{D/2}}
   \int_0^{\infty}d\xi~\xi^{D/2}
    \left(\frac{1}{\tilde g(-\xi^2+i\epsilon)}-\frac{1}{\overline{\tilde g(-\xi^2+i\epsilon)}}\right)J_{\frac{D}{2}-1}(\tau_{xy}\xi)\notag\\
    &=\frac{2(-1)^{1+\frac{D}{2}}\tau_{xy}^{1-\frac{D}{2}}}{(2\pi)^{D/2}}
    \int_0^{\infty}d\xi~\xi^{D/2}
     \frac{\text{Im}\left[\tilde g(-\xi^2+i\epsilon)\right]}{\left|\tilde g(-\xi^2+i\epsilon)\right|^2}J_{\frac{D}{2}-1}(\tau_{xy}\xi).
\end{align}
Using $(x/2)^{1-D/2}J_{\frac{D}{2}-1}(x)\xrightarrow{x\to 0}\Gamma(D/2)^{-1}$ (see e.g. 10.2.2 of \cite{DLMF})
and the fact that $\text{Im}\left[\tilde{g}(-\xi^2+i\epsilon) \right]$
is exponentially damped for large $\xi$ (see Appendix \ref{UVdetEven}),
it can be verified that
\beq
\lim_{x\to y}\rho^{2/D}G(x-y)=\frac{2^{2-\frac{D}{2}}(-1)^{1+\frac{D}{2}}}{(2\pi)^{\frac{D}{2}}\Gamma(\frac{D}{2})}\int_0^{\infty}d \xi~ \xi^{D-1}\frac{\text{Im}\left[\tilde g(-\xi^2+i\epsilon) \right]}{\left|\tilde g(-\xi^2+i\epsilon)\right|^2}.
\eeq
Thus
$G(x-y)$ approaches a constant in the coincidence
limit.\footnote%
{One can understand intuitively why $G_R(x-y)$ is the sum of a
 $\delta$-function with a bounded remainder by noticing that (up to an
 overall numerical factor) our nonlocal d'Alembertian operator has the
 form $1-S$, where the `$1$' corresponds to the first term in \eqref{twopointthree} or \eqref{tatu2}
 and the remainder $S$ is given by an integral-kernel which is both
 bounded and retarded.  The inverse operator $G_R$ would then be
 $G_R=(1+S)^{-1}=1+SG_R=1+S+S^2+S^3\cdots$, a series that should converge
 sufficiently near to $x=y$.  Since the operator $1$ is
 represented by a term of $\delta(x-y)$ in $G_R(x-y)$, one sees that
 $G_R(x-y)$ is the sum of a $\delta$-function with a term involving only
 smooth bounded functions.  }
%
Strictly speaking,
the discussion above
only analyzes the behavior of $G(x-y)$ as $\tau_{xy}$ {\it approaches}
$0$, and consequently it
does not exclude the presence of
terms which blow up discontinuously on the light cone,
such as $\delta(\tau_{xy}^2)$.
However, a similar treatment for the case where $x-y\not=0$ is null
rather than timelike
removes this loophole.



\subsection{A Possible Regularization Scheme for Quantum Field Theoy}
\label{regGCB}
As was shown in the previous Section, changing the usual d'Alambertian
to the nonlocal operator $\Box_{\rho}^{(D)}$ makes the coincidence limit
more divergent, rather than smoothing it out as one might have initially
expected.
But it does so in an interesting way: all the divergences have now been
absorbed into one $\delta$-function at $x=y$.
This feature has a natural application as a regularization tool for
quantum field theory.
In any QFT based on $\Box_{\rho}^{(D)}$,
one would expect
the propagator associated with
internal lines in Feynman diagrams
to
have the same UV behaviour as
\eqref{meh1}.
Subtracting the constant term in \eqref{meh1}
(which corresponds to subtracting a $\delta$-function in real space)
would
then
render all loops finite.
This
would be
a genuinely Lorentzian
regulator,
with no need for
Wick rotation.
It would also be physically motivated,
with the ``UV completion'' being
understood as
a
theory on the causal set.
It would be interesting to
apply this technique to the renormalization of some
well-understood scalar field theories.
\subsection{Stability}
\label{stabGen}
Is the evolution defined by $\Box_{\rho}^{(D)}$  stable?
As we
discussed in Section \ref{org2Dsec},
instabilities
are in general associated with ``unstable modes'', and we agreed to use
this as our criterion of instability for purposes of this paper.
More specifically, we took such a mode to be a plane-wave
$\Phi(x)=e^{ip\cdot x}$ satisfying the equation of motion
$\Box_{\rho}^{(D)}\Phi(x)=0$, with the wave-vector $p$ possessing
a future-directed timelike imaginary part (i.e. $p=p_R+ip_I$ where
$p_I\cdot p_I<0$ and $p_I^0>0$).

The necessary {and} sufficient condition for avoiding unstable modes is
then
\beq
   \tilde{g}(Z)\neq0  \ ,\qquad \forall\ Z\neq0,
\eeq
where $\tilde{g}(Z)$ is defined in \eqref{tildeg}.
Let us argue why this is the case.
First observe that plane solutions of our wave-equation correspond
exactly with zeros of $\tilde{g}(Z)$.
If the above condition is verified,
then the only such zero is at $Z=0$,
just as for the usual d'Alembertian.
But we know (as is also easy to demonstrate ab initio)
that there are no unstable modes in the usual case.
Conversely,
when the above condition is violated for some complex
${Z}\neq0$, it is always possible to find a corresponding
${p}$ with a timelike and future-directed imaginary part which
satisfies
${p}\cdot {p}=\rho^{\frac{2}{D}}Z$.
To see this, we let ${p}={p}_R+i{p}_I$ and take
${p}_R=<\pi^0_R,\vec{\pi}_R>$ and $p_I=<\pi_I,\vec{0}>$ with
$\pi_I>0$. This is always possible because $p_I$ is timelike and
future-directed.
The equations that need to be satisfied are
\beq
  {p}_R\cdot {p}_R-{p}_I\cdot {p}_I=\rho^{\frac{2}{D}}\text{Re}(Z), \qquad
  2{p}_R\cdot {p}_I=\rho^{\frac{2}{D}}\text{Im}(Z). \qquad
\eeq
Substituting for ${p}_I$ leads to
\beq
  \pi^0_R = \frac{\rho^{\frac{2}{D}}\text{Im}({Z})}{-2\pi_I}, \qquad
  |\vec{\pi}_R|^2=\rho^{\frac{2}{D}}\text{Re}({Z})+\frac{\rho^{\frac{4}{D}}\text{Im}({Z})^2}{4\pi_I^2} - \pi_I^2.
\eeq
This system of equations always has a solution.
In fact, there is a whole family of such unstable modes parametrized by $\pi_I$.
Note however that
the condition $|\vec{\pi}_R|^2>0$ puts an upper bound on the value of
$\pi_I$, and therefore on the growth rate of such an instability.

We have thus reduced the question of
whether or not $\Box_{\rho}^{(D)}$ has unstable modes
to the question of whether
$\tilde{g}(Z)$
has zeros other than $Z=0$ in
the complex plane.
We can answer this question by
counting the
zeros of $\tilde{g}(Z)$
with the aid of the
``argument principle"
of complex analysis:
\beq
\frac{1}{2\pi i}\oint\limits_{C}\frac{\tilde{g}'(Z)}{\tilde{g}(Z)}dZ=N-P,
\eeq
where $N$ and $P$ are the number of zeros and poles, respectively,
inside of the closed contour C,
which we choose as shown in Figure \ref{stabilityCont}.
The number of poles inside $C$ is zero because all terms appearing in
$\tilde{g}(Z)$ are finite in that region (at least when $\Lmax$ is finite).
As shown in Figure \ref{stabilityCont}, the path of integration $C$ comprises four pieces:
$C_2$ and $C_4$ run from $-\infty$ to 0 a distance $\epsilon$
above and below the negative real axis respectively,
$C_3$ is a semicircle of radius
$\epsilon$ about the origin, and $C_1$ is (almost) a circle whose radius should be taken to infinity.
For large $Z$ we have from \eqref{uvGen},
\beq
  \frac{\tilde{g}'(Z)}{\tilde{g}(Z)}\xrightarrow{Z\to\infty}\frac{-D2^{D-1}\pi^{\frac{D}{2}-1}\Gamma(D/2)}{2a}b_0Z^{-\frac{D}{2}-1}+\cdots,
\eeq
and it follows that
\beq
  \int\limits_{C_1}\frac{\tilde{g}'(Z)}{\tilde{g}(Z)}dZ=0.
\eeq
(We remind the reader of our standing assumption that $a\neq0$.  See the
remarks following \eqref{meh1}.)
On the other hand
the IR behaviour,
$\tilde{g}(Z)\xrightarrow{Z\to0}-Z$,
leads to
\beq
  \int\limits_{C_3}\frac{\tilde{g}'(Z)}{\tilde{g}(Z)}dZ=i\pi.
\eeq
Also, because $\tilde{g}(x+i\epsilon)=\overline{\tilde{g}(x-i\epsilon)}$ for $x<0$:
\beq
\int\limits_{C_2+C_4}\frac{\tilde{g}'(Z)}{\tilde{g}(Z)}dZ=2i\int\limits_{C_2}\text{Im}\left[\frac{\tilde{g}'(Z)}{\tilde{g}(Z)}\right]dZ.
\eeq
Performing this last integral will allow us to determine whether $\Box_{\rho}^{(D)}$ has unstable modes or not.

Given a choice of the parameters
$a$ and $b_n$,
the last integral can be computed numerically.
In the minimal 4D case discussed in Section \ref{org4Dsec}, we find that
$\Box_{\rho}^{(4)}$ has precisely two ``unstable zeros''.
(Notice that because $\tilde{g}(\bar{Z})=\overline{\tilde{g}(Z)}$, if
$Z$ is a zero of $\tilde{g}(Z)$, so also is $\bar{Z}$.)
We have located these zeros numerically,
as shown in Figure \ref{stability4D}.
With different choices of the parameters $\{a, b_n\}$,
the number of zeros can change, but
we have not been able to find any choice
that would make
$\Box_{\rho}^{(4)}$ stable.
It would be interesting to find an analytical method to check for stability.


\section{Summary and Remarks}
We have defined an infinite family of scalar-field operators on causal
sets which we dubbed Generalized Causet Box (GCB) operators.  For causal
sets made by sprinkling $D$-dimensional Minkowski space $\Minkowski^D$,
these operators reproduce the usual d'Alembertian
$\Box=\nabla_\mu\nabla^\mu$ when
one averages over all sprinklings
and
takes the limit of infinite sprinkling-density $\rho$.
If, on the other hand,
one averages over all sprinklings
while holding $\rho$ fixed,
one obtains an integral operator $\Box_{\rho}^{(D)}$ in $\Minkowski^D$ which is
manifestly Lorentz-invariant, retarded, and nonlocal,
with the degree of nonlocality set by $\rho$.
In the present paper, we have been concerned primarily with these
continuum operators, whose nonlocality can be regarded as a
``mesoscopic'' residue of the underlying causal set discreteness.


The GCB  operators $B^{(D)}_{\rho}$ and their continuum averages $\Box_{\rho}^{(D)}$
are parametrized by a set of coefficients,
and
we derived the equations in these coefficients which ensure
that $\Box$ is recovered in the infrared limit.   The minimal solutions
of these equations turned out to reproduce the original operators
proposed in \cite{Sorkin_2}.
We also computed the
Fourier transform of $\Box_{\rho}^{(D)}$, or equivalently its
``spectrum of eigenvalues''
obtained by applying it to an arbitrary plane wave.
For spacelilke momenta
the spectrum is real.
For timelike momenta
it contains also an imaginary part,
which changes sign under interchange of past with future.
The UV behaviour of the spectrum
differs from that of $\,\Box\,$ in a way which
led
us to propose a genuinely Lorentzian, perturbative regulator for quantum
field theory.

We also studied the question of
whether the evolution defined by the continuum-averaged GCB operators is
stable.
This is of interest in relation to nonlocal field
theories based on $\Box_{\rho}^{(D)}$;
it can also serve as an indicator of the
stability or instability of the corresponding  causet operator $B^{(D)}_{\rho}$.
The continuum-average of the minimal 2D causal set d'Alembertian was shown to be stable by a
direct proof.
In 4D
we did not settle the question analytically, but
we devised a numerical diagnostic that applies to all the operators
$\Box_{\rho}^{(D)}$,
and which disclosed
a pair of unstable modes when applied to the
minimal 4D causal set d'Alembertian.
Are any of the continuum-averaged GCB operators stable in $3+1$
dimensions?  We were not able to find any, but there are an infinite
number of such operators and a definitive search could only be conducted by
analytical means.\footnote
{Also interesting would be an unstable operator whose
 corresponding growth-time was
 either very large (cosmological) or very small (Plankian).
 In the former case, the instability would be irrelevant
 physically, in the latter case it might still be compatible with
 stability of the corresponding discrete evolution.
 We were not able to find any such operator in 4D either. }
Finally, it bears repeating that there might be more reliable indicators
of instability than simply the existence of an exponentially growing
plane-wave solution, which a priori tells us nothing about the behavior
of solutions of limited spatial extent.  For that reason, it would be
worthwhile to analyze directly the late-time behavior of the Green
function $G_R(x-y)$ which is inverse to $\Box_{\rho}^{(D)}$.  If it were
bounded that would imply stability, and if it grew exponentially, that
would imply instability.

Our results also suggest other problems for further work.
It would be interesting, for example,
to work out the continuum-averaged GCB
operators in curved spacetimes.
It was found in \cite{Benincasa_1} that
the minimal 4D operator has the following
limit as $\rho\to\infty$:
$\Box_{\rho}^{(4)}\Phi\to\Box\Phi-\frac{1}{2}R\Phi$, where $R$ is the
Ricci scalar.
(In fact one obtains the same limit in all dimensions $D$.)
Would this feature persist for all of the GCB operators?
This feature has also been used to define an action-functional for
causal sets \cite{Benincasa_1}.
A final question then is whether the instability found
above has any consequences for this causal set action?

\acknowledgments
We thank Niayesh Afshordi for useful discussions throughout the course of this project.

This research was supported in part by NSERC through grant RGPIN-418709-2012.
This research was supported in part by Perimeter Institute for
Theoretical Physics. Research at Perimeter Institute is supported by the
Government of Canada through Industry Canada and by the Province of
Ontario through the Ministry of Research and Innovation.
\bibliographystyle{jhep}
\bibliography{prop}

\newpage
\appendix
\section{IR Behaviour of the GCB Operators: Details}
\label{IRdet}
Here we will derive the equations that the constants $a$ and $\{b_n\}$
should satisfy in order for $\Box^{(D)}_{\rho}$ to have the desired
IR behaviour
\eqref{IR_1}, or equivalently \eqref{IR_2},
which in turn
is equivalent to
\beq
   \tilde{g}(Z) \; \xrightarrow{Z\to0} \;  -Z,
\eeq
where $\tilde{g}(Z)$ is defined by
\beq
  \rho^{-2/D}g^{(D)}_{\rho}(p)\equiv\tilde{g}(Z),
  \label{tildeg2}
\eeq
as given in the right hand side of \eqref{geng}.
\subsection{Even Dimensions}
Let $D=2N+2$ where $N=0,1,2,\dots$.
Then
\beq
  \tilde{g}(Z)=a+2(2\pi)^N\sum_{n=0}^{\Lmax}\frac{b_n}{n!}C_D^n
  \int_{0}^{\infty} s^{2(N+1)n+2N+1}e^{-C_Ds^D}(Z^{1/2}s)^{-N}K_{N}(Z^{1/2}s)\ud s.
  \label{Eventildeg}
\eeq
In order to examine the behaviour of $\tilde{g}(Z)$ as $Z\to0$, we need
to expand $(Z^{1/2}s)^{-N}K_{N}(Z^{1/2}s)$ in this regime.
From
the
power series expansion of $K_N$
(see e.g. 10.31.1 and 10.25.2 of \cite{DLMF}),
it follows that
\begin{subequations}
  \begin{align}
    (Z^{1/2}s)^{-N}K_{N}(Z^{1/2}s)
    =&2^{N-1}(Zs^2)^{-N}\sum_{k=0}^{N-1}\frac{\Gamma(N-k)}{k!}(-Zs^2/4)^{k}\label{exp1e}\\
    &+\frac{(-1)^{N+1}}{2^{N+1}N!}\ln(Z)\label{exp2e}\\
    &+\frac{(-1)^{N}}{2^{N+1}N!}\left[-2\ln(s/2)+\psi(1)+\psi(N+1)\right]\label{exp3e}\\
    &+\frac{(-1)^{N+1}s^2}{2^{N+3}(N+1)!}Z\ln(Z)\label{exp4e}\\
    &+\frac{(-1)^{N}}{2^{N+3}(N+1)!}\left[-2\ln(s/2)+\psi(2)+\psi(N+2)\right]s^2Z\label{exp5e}\\
    &+\mathcal{O}(Z^2), \notag
  \end{align}
\end{subequations}
where $\psi(n)$ is the digamma function.
Because we need the leading behaviour of
$\rho^{-\frac{2}{D}}\tilde{g}(Z)$ to be $-Z$, we have only considered
terms up to this order.
All
the
 terms in \eqref{exp1e} and \eqref{exp2e} diverge as $Z\to0$, forcing
us to pick the $b_n$ such that none of them contribute to $\tilde{g}(Z)$
in the $Z\to0$ limit.
The contribution of the term \eqref{exp4e} is also
unwanted and should be made to vanish by choosing $b_n$ appropriately.
This leads us to the following series of equations:
\beq
  \sum_{n=0}^{\Lmax}\frac{b_n}{n!}C_D^n
  \int_{0}^{\infty} s^{2(N+1)n+2k+1}e^{-C_Ds^D}\ud s=0, \qquad
   k=0,1,\dots,N+1.
\eeq
The integration over $s$ can be performed (see e.g. 5.9.1 of
\cite{DLMF}) to give us the condition reproduced above as equation \eqref{even1}:
\beq
  \sum_{n=0}^{\Lmax}\frac{b_n}{n!}\Gamma(n+\frac{k+1}{N+1})=0, \qquad k=0,1,\dots,N+1.
  \label{evenF1}
\eeq
Requiring the contribution of the constant term \eqref{exp3e} to vanish
yields
\beq
  a+\frac{(-1)^{N+1}2\pi^N}{N!}\sum_{n=0}^{\Lmax}\frac{b_n}{n!}C_D^n
  \int_{0}^{\infty} s^{2(N+1)n+2N+1}e^{-C_Ds^D}\ln(s)\ud s=0.
\eeq
We can perform the integral over $s$ by using the formula (see e.g. 5.9.19 and 5.9.1 of \cite{DLMF})
\beq
  \int_{0}^{\infty}s^{\mu}e^{-as^D}\ln(s)\ud s
  = \frac{\Gamma(\frac{\mu+1}{D})}{D^2a^{\frac{\mu+1}{D}}}\left[\psi(\frac{\mu+1}{D})-\ln(a)\right],
\label{intgLn}
\eeq
leading to \eqref{even2}:
\beq
  a+\frac{2(-1)^{N+1}\pi^N}{D^2C_DN!}\sum_{n=0}^{\Lmax}b_n\psi(n+1)=0.
  \label{evenF2}
\eeq
Finally, requiring the contribution of \eqref{exp5e} to reproduce
the desired $-Z$ behaviour leads to
\beq
  \sum_{n=0}^{\Lmax}\frac{b_n}{n!}C_D^n\int_{0}^{\infty} s^{2(N+1)n+2N+3}e^{-C_Ds^D}\ln(s)\ud s=\frac{2(-1)^N(N+1)!}{\pi^N}.
\eeq
Performing the integral using \eqref{intgLn}
furnishes
 \eqref{even3}:
\beq
\sum_{n=0}^{\Lmax}\frac{b_n}{n!}\Gamma(n+\frac{N+2}{N+1})\psi(n+\frac{N+2}{N+1})=\frac{2(-1)^N(N+1)!}{\pi^N}D^2C_D^{\frac{N+2}{N+1}}.
\label{evenF3}
\eeq
\subsection{Odd Dimensions}
Let $D=2N+1$ where $N=0,1,2,\dots$.
Then
\beq
  \tilde{g}(Z)=a+2(2\pi)^{N-1/2}\sum_{n=0}^{\Lmax}\frac{b_n}{n!}C_D^n
\int_{0}^{\infty} s^{2(N+1)n+2N}e^{-C_Ds^D}(Z^{1/2}s)^{-N+1/2}K_{N-1/2}(Z^{1/2}s)\ud s.
 \label{Oddtildeg}
\eeq
From
the power series expansion of $K_N$ (see $10.27.4$  of and $10.25.2$ of \cite{DLMF}),
it follows that
\begin{subequations}
  \begin{align}
    (Z^{\frac{1}{2}}s)^{-N+\frac{1}{2}}K_{N-\frac{1}{2}}(Z^{\frac{1}{2}}s)
    =&(-1)^{N-1}2^{N-\frac{3}{2}}\pi(Z^{\frac{1}{2}}s)^{-2N+1}
    \sum_{k=0}^{N}\frac{(Zs^2/4)^k}{k!\Gamma(k-N+\frac{3}{2})}\label{exp1o}\\
    &+\frac{(-1)^N2^{-N-\frac{1}{2}}\pi}{\Gamma(N+\frac{1}{2})}\label{exp2o}\\
    &+\frac{(-1)^N2^{-N-\frac{5}{2}}\pi}{\Gamma(N+\frac{3}{2})}s^2Z\label{exp3o}\\
    &+\mathcal{O}(Z^{3/2}). \notag
  \end{align}
\end{subequations}
As before, we have only kept track of terms up to $Z$.
The contributions of all the terms in \eqref{exp1o} should be made to
vanish; this leads to the equation
\beq
    \sum_{n=0}^{\Lmax}\frac{b_n}{n!}C_D^n
    \int_{0}^{\infty} s^{2(N+1)n+2k+1}e^{-C_Ds^D}\ud s=0, \qquad k=0,1,\dots,N.
\eeq
Performing the integral over $s$ gives us \eqref{odd1}:
\beq
  \sum_{n=0}^{\Lmax}\frac{b_n}{n!}\Gamma(n+\frac{2k+2}{2N+1})=0, \qquad k=0,1,\dots,N.
\label{oddF1}
\eeq
Requiring the contribution of the constant term \eqref{exp2o} to vanish yields
\beq
   a+\frac{(-1)^{N}\pi^{N+\frac{1}{2}}}{\Gamma(N+\frac{1}{2})}\sum_{n=0}^{\Lmax}\frac{b_n}{n!}C_D^n
   \int_{0}^{\infty} s^{2(N+1)n+2N}e^{-C_Ds^D}\ud s=0 \ ,
\eeq
which is equivalent to \eqref{odd2}:
\beq
  a+\frac{(-1)^{N}\pi^{N+\frac{1}{2}}}{DC_D\Gamma(N+\frac{1}{2})}\sum_{n=0}^{\Lmax}b_n=0
  \ .
\label{oddF2}
\eeq
Finally, requiring the contribution of \eqref{exp3o} to reproduce the
desired $-Z$ behaviour leads to
\beq
  \frac{(-1)^{N}\pi^{N+\frac{1}{2}}}{4\Gamma(N+\frac{3}{2})}\sum_{n=0}^{\Lmax}\frac{b_n}{n!}C_D^n
  \int_{0}^{\infty} s^{2(N+1)n+2N+2}e^{-C_Ds^D}\ud s=-1,
\eeq
which
furnishes
\eqref{odd3}:
\beq
  \sum_{n=0}^{\Lmax}\frac{b_n}{n!}\Gamma(n+\frac{2N+3}{2N+1})
  = \frac{(-1)^{N-1}4\Gamma(N+\frac{3}{2})}{\pi^{N+\frac{1}{2}}}DC_D^{\frac{2N+3}{2N+1}}.
\label{oddF3}
\eeq
\section{UV Behaviour of the GCB Operators: Details}
\label{UVdet}
Here we derive the UV behaviour of $\Box^{(D)}_{\rho}$.
We will make use of the following identity \cite{LTRLI},
which holds for arbitrary natural number $m$:
\beq
  \frac{K_{p}(Z^{\frac{1}{2}}s)}{Z^{\frac{p}{2}}}
  =(-1)^m\left(\frac{2}{s}\right)^m\frac{d^m}{dZ^m}\left\{\frac{K_{p-m}(Z^{\frac{1}{2}}s)}{Z^{\frac{p-m}{2}}}\right\}.
  \label{UVtrick}
\eeq
\subsection{Even Dimensions}\label{UVdetEven}
Let $D=2N+2$ where $N=0,1,2,\dots$, and $p=m=N$ in \eqref{UVtrick}. It then follows that
\beq
   Z^{-N/2}K_N(Z^{\frac{1}{2}}s)=(-1)^N\left(\frac{2}{s}\right)^N\frac{d^N}{dZ^N}K_{0}(Z^{\frac{1}{2}}s).
\eeq
Substituting this in the definition of $\tilde{g}(Z)$, as given by \eqref{Eventildeg}, produces
\beq
  \tilde{g}(Z)=a+(-1)^N2^{2N+1}\pi^N\sum_{n=0}^{\Lmax}\frac{b_n}{n!}C_D^n
  \frac{d^N}{dZ^N}I^{(D)}_n(Z) \ ,
  \label{gTildeEven}
\eeq
where
\beq
  I^{(D)}_n(Z)\equiv\int_{0}^{\infty} s^{Dn+1}e^{-C_Ds^D}K_{0}(Z^{\frac{1}{2}}s)\ud s \ .
  \label{ieven}
\eeq
It then suffices to study the behaviour of
this integral
as $Z\to\infty$.
It follows from $10.29.4$ of \cite{DLMF} that
\beq
  K_{0}(Z^{\frac{1}{2}}s)=\frac{-1}{Z^{\frac{1}{2}}s}\frac{d}{ds}\left(sK_{1}(Z^{\frac{1}{2}}s)\right).
\eeq
Plugging this relation in \eqref{ieven} and integrating by parts yields
\beq
   I^{(D)}_n(Z)
   =-\frac{1}{Z^{\frac{1}{2}}}\left\{s^{Dn+1}e^{-C_Ds^D}K_{1}(Z^{\frac{1}{2}}s)
   \Big|_{0}^{\infty}-\int_{0}^{\infty}sK_{1}(Z^{\frac{1}{2}}s)\frac{d}{ds}(s^{Dn}e^{-C_Ds^D})\ud s\right\}.
\eeq
The first term vanishes when evaluated at $\infty$.  When evaluated at
$0$, it is non-zero only when $n=0$, because
$K_{1}(Z^{\frac{1}{2}}s)\to Z^{\frac{1}{2}}s^{-1}$
when $s\to0$.
It then follows that
\beq
  I^{(D)}_n(Z)=\frac{1}{Z^{\frac{1}{2}}}
  \left\{\frac{\delta_{n0}}{Z^{\frac{1}{2}}}+\int_{0}^{\infty}sK_{1}(Z^{\frac{1}{2}}s)\frac{d}{ds}(s^{Dn}e^{-C_Ds^D})\ud s\right\}.
  \label{ieven2}
\eeq
From
$10.29.3$ of \cite{DLMF},
\beq
  K_{1}(Z^{\frac{1}{2}}s)=\frac{-1}{Z^{\frac{1}{2}}}\frac{d}{ds}K_{0}(Z^{\frac{1}{2}}s).
\eeq
Plugging this back into \eqref{ieven2} and integrating once
again by parts
yields
\beq
  I^{(D)}_n(Z)=\frac{1}{Z}\left\{\delta_{n0}
  +\int_{0}^{\infty}K_{0}(Z^{\frac{1}{2}}s)\frac{d}{ds}\left[s\frac{d}{ds}(s^{Dn}e^{-C_Ds^D})\right]\ud s\right\}.
  \label{ieven3}
\eeq
It can be shown that
\beq
\lim_{Z\to\infty}\int_{0}^{\infty}K_{0}(Z^{\frac{1}{2}}s)\frac{d}{ds}\left[s\frac{d}{ds}(s^{Dn}e^{-C_Ds^D})\right]\ud s=0.
\eeq
With the aid of
\eqref{gTildeEven},
it then follows that for large $Z$,
\beq
   \tilde{g}(Z)
   =
  a \;+\; 2^{D-1}\pi^{\frac{D}{2}-1}\Gamma(D/2)b_0 \, Z^{-\frac{D}{2}} \;+\cdots
\eeq
Notice that both these terms are real for both positive and negative
$Z$, because $D/2$ is an integer when $D$ is even.
In order to produce the sub-leading terms,
one can continue
integrating by parts
in \eqref{ieven3}.
The
sub-leading terms
are thus also real,
whence
the imaginary part of $\tilde g(Z)$
must,
for even $D$,
decay faster than
any power of $Z$ for $Z\to\infty$.
This behavior can be seen in Figures \ref{g2Dfig} and \ref{g4Dfig}.
\subsection{Odd Dimensions}
Let $D=2N+1$ where $N=0,1,2,\dots$, and $p=m-\frac{1}{2}=N-\frac{1}{2}$ in \eqref{UVtrick}. It then follows that
\beq
   Z^{\frac{1-2N}{4}}K_{N-\frac{1}{2}}(Z^{\frac{1}{2}}s)=(-1)^N\left(\frac{2}{s}\right)^N\frac{d^N}{dZ^N}\{Z^{\frac{1}{4}}K_{-\frac{1}{2}}(Z^{\frac{1}{2}}s)\}.
\eeq
From  $10.39.2$ of \cite{DLMF}, we have that
\beq
   K_{-\frac{1}{2}}(Z^{\frac{1}{2}}s)=Z^{-\frac{1}{4}}\left(\frac{\pi}{2s}\right)^{\frac{1}{2}}e^{-Z^{\frac{1}{2}}s}\ ,
\eeq
whence
\beq
  Z^{\frac{1-2N}{4}}K_{N-\frac{1}{2}}(Z^{\frac{1}{2}}s)=\frac{(-1)^N2^{N-\frac{1}{2}}\pi^{\frac{1}{2}}}{s^{N+\frac{1}{2}}}\frac{d^N}{dZ^N}e^{-Z^{\frac{1}{2}}s}.
\eeq
Substituting this into the definition of $\tilde{g}(Z)$, as given by \eqref{Oddtildeg}, produces
\beq
  \tilde{g}(Z)=a+(-1)^N2^{2N}\pi^N\sum_{n=0}^{\Lmax}\frac{b_n}{n!}C_D^n
  \frac{d^N}{dZ^N}I^{(D)}_n(Z),
  \label{gTildeOdd}
\eeq
where
\beq
   I^{(D)}_n(Z)\equiv\int_{0}^{\infty} s^{Dn}e^{-C_Ds^D}e^{-Z^{\frac{1}{2}}s}\ud s.
   \label{iodd}
\eeq
It then suffices to study the behaviour of
this integral
as $Z\to\infty$:
\begin{align}
   I^{(D)}_n(Z)&=-Z^{-\frac{1}{2}}\int_{0}^{\infty}s^{Dn}e^{-C_Ds^D}\frac{d}{ds}e^{-Z^{\frac{1}{2}}s}\ud s\notag\\
   &=-Z^{-\frac{1}{2}}\left\{s^{Dn}e^{-C_Ds^D-Z^{\frac{1}{2}}s}\Big|_0^{\infty}-\int_{0}^{\infty}e^{-Z^{\frac{1}{2}}s}\frac{d}{ds}(s^{Dn}e^{-C_Ds^D})\right\}\notag\\
   &=Z^{-\frac{1}{2}}\left\{\delta_{n0}+\int_{0}^{\infty}e^{-Z^{\frac{1}{2}}s}\frac{d}{ds}(s^{Dn}e^{-C_Ds^D})\right\}.
\end{align}
Again, because
\beq
  \lim_{Z\to\infty}\int_{0}^{\infty}e^{-Z^{\frac{1}{2}}s}\frac{d}{ds}(s^{Dn}e^{-C_Ds^D})=0,
\eeq
we can deduce from \eqref{gTildeOdd} that
\beq
   \tilde{g}(Z) \;=\; a \;+\; 2^{D-1}\pi^{\frac{D}{2}-1}\Gamma(D/2)b_0Z^{-\frac{D}{2}}\; +\cdots \ .
\eeq
\section{Derivation of Equation \eqref{g2Dexact}}
\label{deriveExact2D}
From the general equations,  \eqref{gGen} and \eqref{chiP}, we have
\beq
  \rho^{-1}g^{(2)}_{\rho}(p)=a^{(2)}+\rho\sum_{n=0}^{2}\frac{(-1)^n\rho^n}{n!}b^{(2)}_n\frac{\partial^n}{\partial\rho^n}\chi(p,\rho),
  \label{exp2D}
\eeq
where $\{a^{(2)},b_n^{(2)}\}$ are given in \eqref{coef2Dmin} and
\beq
  \chi(p,\rho)=2\int_0^{\infty}se^{-\rho s^2/2}K_0(\sqrt{\Lp}s)\ud s.
\eeq
From
 the relation (see e.g. 8.6.6 and 8.19.1 of \cite{DLMF}),
\beq
  e^ZE_1(Z)=2\int_0^{\infty}e^{-t}K_0(\sqrt{2zt})\ud t,
\eeq
it follows that
\beq
  \chi(p,\rho)=\rho^{-1}e^{Z/2}E_1(Z/2)  \ , \qquad
  Z=\rho^{-1}\Lp.
\eeq
Furthermore, using the identities (see e.g. 8.9.14 and 8.19.12 of \cite{DLMF}),
\begin{align}
   \frac{d}{dz}\left[e^{z}E_p(z)\right]&=e^zE_p(z)\left(1+\frac{p-1}{z}\right)-\frac{1}{z},\\
   pE_{p+1}(z)+zE_{p}(z)&=e^{-z}\label{ident},
\end{align}
it can be shown that
\begin{align}
  \rho^2\frac{\partial\chi}{\partial\rho}&=e^{Z/2}E_2(Z/2)-e^{Z/2}E_1(Z/2)\\
  \rho^3\frac{\partial^2\chi}{\partial\rho^2}&=e^{Z/2}E_1(Z/2)\left[2+Z/2\right]-e^{Z/2}E_2(Z/2)\left[3+Z/2\right].
\end{align}
Equation \eqref{g2Dexact} results
from
plugging these expressions back into \eqref{exp2D} and using \eqref{ident}:
\beq
\rho^{-1}g^{(2)}_{\rho}(p)=-Ze^{Z/2}\text{E}_2(Z/2).
\eeq
\section{Damping the fluctuations}
\label{dampFluct}
In reference \cite{Sorkin_2}
a prescription was given
to get from the causet d'Alembertian $B^{(2)}_{\rho}$
of \eqref{min2Ddis}
a new operator
$\tilde{B}^{(2)}_{\rho,\epsilon}$,
whose fluctuations are damped, but
which has the same mean over
sprinklings
as $B^{(2)}_{\tilde\rho}$
with $\tilde{\rho}=\epsilon\rho$.
Here we
generalize this
prescription to the class of causet d'Alembertians $B^{(D)}_{\rho}$
defined in \eqref{genDisBox}.
(See Sections \ref{org2D4D} and \ref{GenCB} for any
symbol
which is not defined in what follows.)

Given the causal set d'Alembertian,
\beq
  \rho^{-2/D}(B^{(D)}_{\rho}\Phi)(x)=a\Phi(x)+\sum_{m=0}^{\Lmax}b_m\sum_{y \in I_m}\Phi(y),
\eeq
we construct as follows a new operator
$\tilde{B}^{(D)}_{{\rho},\epsilon}$
whose effective non-locality energy-density scale is $\epsilon{\rho}$:
\beq
  \tilde{\rho}^{-2/D}(\tilde{B}^{(D)}_{\rho,\epsilon}\Phi)(x)=a\Phi(x)+\sum_{n=0}^{\infty}\tilde{b}_n\sum_{y \in I_n}\Phi(y),
\eeq
with
\beq
  \tilde{b}_n=\epsilon(1-\epsilon)^n\sum_{m=0}^{\Lmax}\binom{n}{m}\frac{b_m\epsilon^m}{(1-\epsilon)^m}, \qquad
  \epsilon=\tilde{\rho}/\rho.
\eeq
(Here, the binomial coefficient $\binom{n}{m}$ is zero by convention for $m>n$.)

Let us demonstrate that the continuum limit of
$\tilde{B}^{(D)}_{{\rho},\epsilon}$,
which we will denote by
$\tilde{\Box}^{(D)}_{\tilde{\rho}}$, is equal to
$\Box^{(D)}_{\tilde{\rho}}$:
\begin{subequations}
        \begin{align}
                &\tilde{\rho}^{-2/D}(\tilde{\Box}^{(D)}_{\tilde{\rho}}\Phi)(x)-a\Phi(x)\notag\\
                &=\rho\sum_{n=0}^{\infty}\frac{\tilde{b}_n}{n!}\int\limits_{J^{-}(x)}e^{-\rho V(x,y)}[\rho V(x,y)]^n\phi(y)dV_y\notag\\
                &=\rho\epsilon\sum_{m=0}^{\Lmax}\frac{b_m\epsilon^m}{m!}\int\limits_{J^{-}(x)}e^{-\rho V(x,y)}\left\{\sum_{n=m}^{\infty}\frac{(1-\epsilon)^{n-m}}{(n-m)!}\left[\rho V(x,y)\right]^n\right\}\phi(y)dV_y\notag\\
                &=\tilde{\rho}\sum_{m=0}^{\Lmax}\frac{b_m}{m!}\int\limits_{J^{-}(x)}e^{-\rho V(x,y)}\left\{\sum_{n=m}^{\infty}\frac{(1-\epsilon)^{n-m}}{(n-m)!}\left[\rho V(x,y)\right]^{n-m}\right\}\left[\epsilon\rho V(x,y)\right]^{m}\phi(y)dV_y\notag\\
                &=\tilde{\rho}\sum_{m=0}^{\Lmax}\frac{b_m}{m!}\int\limits_{J^{-}(x)}e^{-\rho V(x,y)}e^{(1-\epsilon)\rho V(x,y)}\left[\tilde{\rho} V(x,y)\right]^{m}\phi(y)dV_y\notag\\
                &=\tilde{\rho}\sum_{m=0}^{\Lmax}\frac{b_m}{m!}\int\limits_{J^{-}(x)}e^{-\tilde{\rho} V(x,y)}\left[\tilde{\rho} V(x,y)\right]^{m}\phi(y)dV_y.\notag\\
                &=\tilde{\rho}^{-2/D}(\Box^{(D)}_{\tilde{\rho}}\Phi)(x)-a\Phi(x).\notag
        \end{align}
\end{subequations}

Of course, we have not proven here that the fluctuations of
$\tilde{B}^{(D)}_{\tilde{\rho}}$ are actually damped.
This has been confirmed numerically for the minimal 2D and 4D operators
in \cite{Sorkin_2} and \cite{Benincasa_1}.
It would be interesting to confirm
it also
for the full set of GCB operators in all dimensions.



\end{document}